\documentclass[journal,10pt]{IEEEtran}
\usepackage[T1]{fontenc}
\usepackage{float}
\usepackage{dblfloatfix}
\usepackage[dvips]{graphicx}
\graphicspath{{../eps/}}
\DeclareGraphicsExtensions{.eps}
\ifCLASSINFOpdf
\else
\fi

\ifCLASSOPTIONcompsoc
  \usepackage[caption=false,font=normalsize,labelfont=sf,textfont=sf]{subfig}
\else
  \usepackage[caption=false,font=footnotesize]{subfig}
\fi

\usepackage{amsmath}
\usepackage[cmintegrals]{newtxmath}
\usepackage{graphicx}
\usepackage{xcolor}
\usepackage{url}
\newtheorem{theorem}{Theorem}
\newtheorem{proposition}{Proposition}
\newtheorem{corollary}{Corollary}
\newtheorem{observation}{Observation}
\newtheorem{remark}{Remark}
\begin{document}

\title{On the Outage Analysis and Finite SNR Diversity-Multiplexing Tradeoff of Hybrid-Duplex Systems for
Aeronautical Communications}

\author{Tan~Zheng~Hui~Ernest,
				A~S~Madhukumar,
				Rajendra~Prasad~Sirigina,
				and~Anoop~Kumar~Krishna%
\thanks{Tan Zheng Hui Ernest is with the School of Computer Science and Engineering, Nanyang Technological University, Singapore e-mail: (tanz0119@e.ntu.edu.sg).}
\thanks{A S Madhukumar is with the School of Computer Science and Engineering, Nanyang Technological University, Singapore e-mail: (asmadhukumar@ntu.edu.sg).}
\thanks{Rajendra Prasad Sirigina is with the School of Computer Science and Engineering, Nanyang Technological University, Singapore e-mail: (raje0015@ntu.edu.sg).}
\thanks{Anoop Kumar Krishna is with Airbus Singapore Pte Ltd, Singapore e-mail: (anoopkumar.krishna@airbus.com).}}

\maketitle

\begin{abstract}
A hybrid-duplex aeronautical communication system (HBD-ACS) consisting of a full-duplex (FD) enabled ground station (GS), and two half-duplex (HD) air-stations (ASs) is proposed as a direct solution to the spectrum crunch faced by the aviation industry. Closed-form outage probability and finite signal-to-noise ratio (SNR) diversity gain expressions in aeronautical communications over Rician fading channels are derived for a successive interference cancellation (SIC) detector. Similar expressions are also presented for an interference ignorant (II) detector and HD-equivalent modes at GS and ASs. Through outage and finite SNR diversity gain analysis conducted at the nodes, and system level, residual self-interference (SI) and inter-AS interference are found to be the primary limiting factors in the proposed HBD-ACS. Further investigations revealed that the II and SIC detectors in the proposed HBD-ACS are suitable for weak and strong interference scenarios, respectively. When compared to HD-ACS, the proposed HBD-ACS achieves lower outage probability and higher diversity gains at higher multiplexing gains when operating at low SNRs. Finite SNR analysis also showed the possibility of the proposed HBD-ACS being able to attain interference-free diversity gains through proper management of residual SI. Hence, the proposed HBD-ACS is more reliable and can provide better throughput compared to existing HD-ACS at low-to-moderate SNRs.
\end{abstract}


\IEEEpeerreviewmaketitle

\section{Introduction}

\IEEEPARstart{B}{etween} 2012 and 2032, air travel within the Pacific South East Asia region is projected to record a compounded annual growth rate of 5.3\% \cite{icao2016long}. This air travel growth trend exposes existing aeronautical communication systems (ACSs) to considerable strain due to demand for data communications from legacy, current and future generation avionics systems. Consequently, this places an additional strain on existing Air-to-Ground (A/G) and Air-to-Air (A/A) aeronautical communication links on the congested aeronautical spectrum. With existing ACSs being unable to deliver the needed data capacity \cite{neji2013survey}, various communication technologies have been proposed to improve the capabilities of existing A/G and A/A links \cite{neji2013survey,ernest2016efficiency}. However, these solutions do not directly address the issue of spectrum utilization.

A hybrid-duplex (HBD) ACS consisting of half-duplex (HD) air-stations (ASs) operating existing avionics systems with full-duplex (FD) ground stations (GSs) can be an alternative solution to the shortage of available aeronautical spectrum currently faced by the aviation community. Changes to existing/legacy HD avionics systems currently on board aircrafts can be kept to a minimum in HBD-ACS, thus enabling HBD-ACS to be less disruptive to adopt. Wireless communication systems that have adopted the HBD paradigm include cognitive radio systems \cite{li2014linear} and cellular systems \cite{mohammadi2015full, jang2015spatial, cirik2018robust}. 

In HBD systems, both FD and HD nodes communicate on the same spectrum since an FD node can simultaneously transmit and receive signals on the same frequency and thereby improve the spectral efficiency \cite{kim2015survey}, \cite{sabharwal2014band}, \cite{tapio2013system}. Despite extensive studies done on self-interference (SI) mitigation architectures, SI remains the primary challenge faced by FD nodes due to simultaneous signal transmission and reception. SI mitigation architectures can be categorized into either passive suppression or active cancellation \cite{sahai2013impact}. The former mitigates SI through induced path loss (e.g. antenna separation) while the latter cancels SI in the analog or digital domain. However, residual SI will still be present due to the limited dynamic range of analog-to-digital converters  \cite{sabharwal2014band}, \cite{korpi2014full}, inherent carrier phase noise \cite{syrjala2014analysis} and imperfect SI channel estimation at the FD transceiver \cite{sahai2013impact}. Effectively managing residual SI in HBD-ACS opens up the possibility of directly addressing the spectrum crunch faced by the aviation industry. In particular, multiple aircrafts and ground stations can communicate on the same aeronautical spectrum, providing motivation for this paper. 

\subsection{Related Literature}
Apart from SI at FD nodes, HD nodes in HBD systems also experience interference due to transmissions from other HD and FD nodes. In the literature, multiple interference management approaches have been presented. However, this paper focuses on two widely known approaches where interference is either ignored, i.e., interference ignorant (II) detector, or successfully canceled, i.e., successive interference cancellation (SIC) detector. 

To quantify the effectiveness of the II and SIC detectors, many related works in literature have attempted to determine the closed-form outage probabilities of these detectors under various fading models. Having a closed-form outage probability expression enables a system's packet error rate, i.e., link availability, to be analyzed if the transmitted signals span over one fading block \cite{lin2013outage}. For the II detector, closed-form outage expressions for Nakagami-$m$ fading \cite{yao1992investigations} and composite fading consisting of exponentially distributed signal-of-interest (SOI) and squared ${\mathcal{K}}$-distributed interfering signals \cite{bithas2015mobile} have been noted. It should be pointed out that \cite{yao1992investigations} and \cite{bithas2015mobile} are only applicable to specific fading environments and may not be applicable for all aeronautical scenarios where Rician fading is experienced. To this end, a recent paper by Rached et al. \cite{rached2017unified} presented generalized outage probability expressions that apply to a wide variety of fading scenarios, including Rician fading.

Multiple works on outage expressions for SIC detectors have been noted. For instance, SIC outage expressions were investigated by Hasna et al. \cite{hasna2003performance} and Romero-Jerez and Goldsmith \cite{romero2008receive}, but these studies only considered partial SIC where at least one interfering signal remains after interference cancellation. A closed-form outage expression for SIC was studied by Weber et al. \cite{weber2007transmission} for nodes distributed via a Poisson point process. The work in \cite{weber2007transmission} did not consider fading and receiver noise in the signal model, and thus, the closed-form expressions are not directly applicable for aeronautical communications. A recent paper by Zhang et al. \cite{zhang2017full} presented outage probability expressions for a two-stage SIC detector. However, the outage expressions are specific for Rayleigh fading scenarios and are not applicable to Rician fading scenarios that are common in aeronautical communications. From the mentioned studies, hitherto closed-form outage probability expressions for SIC detectors in Rician fading aeronautical scenarios remain an open problem.

Apart from outage probability, both finite signal-to-noise ratio (SNR) diversity gain and finite SNR diversity-multiplexing trade-off (DMT) are metrics that can be used to measure the effectiveness of II or SIC detectors in fixed and variable transmission rate systems, respectively. In particular, both finite SNR diversity gain and finite SNR DMT quantifies the slope of outage probability curves at a particular SNR \cite{shin2008diversity}, with the latter considering multiplexing gain \cite{narasimhan2006finite}. Finite SNR analysis can reveal outage deviation behaviors, which are not present at asymptotically high SNRs due to fading statistics \cite{shin2008diversity}. From a practical perspective, analyzing outage probability decay rates, i.e., finite SNR diversity gain, provides an accurate picture of a system's outage performance since wireless communication systems are typically designed to operate at low-to-moderate SNR ranges. It has also been pointed out by Narasimhan \cite{narasimhan2006finite} that finite SNR diversity gain analysis can be used to estimate the SNR needed to achieve a particular rate of error decay, which can be done through turbo codes or low-density parity-check codes. More crucially, outage probability and diversity gain can be used to gauge the upper and lower limits of a system's bit error rate performance \cite{zheng2003diversity,nabar2005diversity}. 

Finite SNR analysis for Nakagami-$m$ \cite{wang2012finite} and Rayleigh fading \cite{lin2013outage,yang2015efficient} scenarios have also been studied. However, the conclusions drawn in these studies are specific to Nakagami-$m$ and Rayleigh fading and are not fully applicable for ACS since Rician fading scenarios, typically encountered by ACS, are not considered. Studies on finite SNR analysis for Rician fading channels have been seen. The impact of Rician $K$ factors on outage behavior and finite SNR DMT for multiple-input multiple-output (MIMO) systems was investigated by Narasimhan \cite{narasimhan2006finite} and Shin et al. \cite{shin2008diversity}. A recent paper by Heidarpour et al. \cite{heidarpour2017finite} saw finite SNR DMT analysis being applied to analyze the performance of a network coded cooperative communication system. Despite the noted studies, there is still room for further work on finite SNR DMT analysis for HBD-ACS. 

\subsection{Main Contributions}
The main contributions of this paper are summarized as follows:

\begin{itemize}
	\item This paper presents closed-form expressions for outage probability, finite SNR diversity gain, and finite SNR DMT for a II detector and a two-stage SIC detector in a Rician fading environment.
	\item It is shown that the proposed HBD-ACS attains superior outage performance over existing HD-ACS at low SNRs. At high SNRs, however, the outage performance of the proposed HBD-ACS is eclipsed by HD-ACS as the former becomes interference-limited. Nonetheless, we show through numerical simulations that the HBD-ACS can meet typical Quality-of-Service (QoS) requirements, e.g., frame error rate $\leq 10^{-3}$, at high SNRs for a range of interference levels through II and SIC detectors. 
	\item Unlike \cite{etkin2008gaussian} and \cite{sirigina2016symbol}, the desired and interfering signal levels are related through a scaling parameter. In contrast to the results in \cite{sirigina2016symbol}, it is shown that the asymptotic diversity gain of the SIC detector is zero for all interference levels.
	\item The HD-ACS is shown to achieve better diversity gain than the proposed HBD-ACS at low multiplexing gains. However, at high multiplexing gains, the HD-ACS achieves zero diversity gain while the proposed HBD-ACS achieves non-zero diversity gain.
\end{itemize}

\subsection{Relevance to Related Literature}
In this work, full interference cancellation is assumed for the two-stage SIC detector. This is unlike in \cite{hasna2003performance} and \cite{romero2008receive} where only partial SIC is assumed. In addition, the impact of interference on the proposed HBD-ACS is analyzed from the outage probability and finite SNR DMT perspective, which was not covered in \cite{li2014linear} - \cite{cirik2018robust}, \cite{yao1992investigations} - \cite{rached2017unified}, \cite{weber2007transmission} and \cite{zhang2017full}. In contrast to \cite{shin2008diversity}, \cite{narasimhan2006finite} and \cite{heidarpour2017finite}, this work extends upon the outage and finite SNR DMT analysis framework to jointly identify interference scenarios for the proposed single-input-single-output equivalent HBD-ACS.

The remainder of this paper is organized as follows. The system model is introduced in Section II, with closed-form outage probability expressions at GS and AS-2 presented in Section III. In Section IV, finite SNR diversity gain expressions for both HBD-ACS and HD-ACS are derived and analyzed. Numerical results are then presented in Section V before the conclusion of the paper in Section VI.

\section{System Model}
\begin{figure} [tpb]
\centering
\includegraphics [width=0.8\columnwidth]{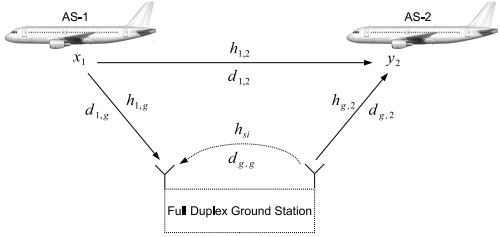} 
\vspace{-2cm}
\caption{Air-Station 1 (AS-1) and Air-Station 2 (AS-2) operating in HD mode while communicating with the FD ground station (GS).}
\label{fig:1}
\end{figure}

In this paper, A/G communications involving an FD-enabled GS node with two HD ASs in an A/G link is studied. Specifically, a scenario with Air-Station 1 (AS-1) transmitting signals to the GS while Air-Station 2 (AS-2) is receiving signals from the GS is assumed. For the HD transceivers at AS-1 and AS-2, a single-antenna configuration with separate transmit and receive radio frequency chains is assumed. In contrast, the FD transceiver at the GS is assumed to be configured with one transmit antenna and one receive antenna through separate radio frequency chains. Due to the fact that the GS node is FD-capable, the HD AS-1 and HD AS-2 simultaneously transmits and receives, respectively, signals on the same aeronautical spectrum (e.g. VHF, L-band) as the GS. Therefore, AS-1 interferes with communications at AS-2 when the latter receives signals from GS.

In this work, an SI mitigation architecture with a shared local oscillator is assumed at the FD-enabled GS. Such a setup enables lower levels of phase noise to be experienced \cite{syrjala2014analysis}, \cite{li2018self}. As such, we only consider residual SI at the FD-enabled GS as a result of imperfect SI channel estimation and phase noise \cite{sahai2013impact}. Furthermore, the SI link ($h_{si}$) is modeled as a Rician fading channel to account for passive and active SI mitigation \cite{ahmed2015all} .\footnote{Depending on the degree of passive and active SI mitigation, the resultant SI channel ($h_{si}$) can be a Rician or Rayleigh fading channel \cite{ahmed2015all}. Thus, modeling $h_{si}$ as a Rician fading channel enables the degree of passive and active SI mitigation to be defined through the Rician $K$ factor.} Thus, an II detector is considered at the FD-enabled GS since signal detection is performed in the presence of residual SI.

Rician fading aeronautical communications channels in an en route scenario is assumed to provide a realistic evaluation of the HBD-ACS \cite{haas2002aeronautical,matolak2017air_suburban,yuan2018capacity}. Following the work in \cite{haas2002aeronautical} and \cite{yuan2018capacity}, the link between AS-1 and AS-2 is also modeled as a Rician fading channel. Accordingly, we assume that the ASs are communicating with the GS at cruising altitude, with the signal model of this work based on \cite{sahai2013impact}. Also, the effect of Doppler shift is assumed to be compensated in this work \cite{lee2018uav}. \footnote{It is useful to note that Doppler shift is not a performance limitation in the upcoming L-band digital aeronautical communication systems (LDACS) standard \cite{jamal2017fbmc}.}

\subsection{Ground Station}

Let $x_1[t]$ and $x_{gs}[t]$ be the signals transmitted by AS-1 and GS, respectively, and $h_{1,g}[t]$ be the channel from AS-1 to GS. Additionally, let $x_{si}[t]$ be the SI signal at GS and let $h_{si}$ be the SI channel gain. From the perspective of GS, $x_{si}[t]=x_{gs}[t]$. The received signal at GS can be written as 
\begin{eqnarray} \label{y_gs}
y_{gs}[t] & = & \sqrt{\Omega_{X}}h_{1,g}[t]x_{1}[t] + \sqrt{\Omega_X\alpha_{g,g}} \cdot |\widetilde{h}_{si}|x_{si}[t] \nonumber \\
 & & + \sqrt{\Omega_X\alpha_{g,g}}|h_{si}|\gamma_{\phi}w_{\phi}[t] + w_{g}[t],
\end{eqnarray}
where $\widetilde{h}_{si}$ is the error of the imperfect SI channel gain estimate, defined as $\widetilde{h}_{si}=h_{si}-\widehat{h}_{si}$, and $\widehat{h}_{si}$ is the imperfect estimation of the SI channel gain. In addition, let $\widetilde{h}_{si}$ be modeled as a zero mean, circularly symmetric complex Gaussian random variable (RV) with variance $\epsilon$ to quantify the SI channel estimation error \cite{zlatanov2017capacity}. Modeling $\widetilde{h}_{si}$ as a zero mean Gaussian RV with variance $\epsilon$ enables the system to model the worst case residual SI \cite{zlatanov2017capacity}. Also, let $w_{g}[t]$ be the GS additive white Gaussian noise (AWGN) with zero mean and variance $\sigma^2_{g}$, and let the phase noise term $w_{\phi}[t]$ follow a Gaussian distribution with zero mean and unit variance, scaled by the scaling factor $\gamma_{\phi}$ \cite{sahai2013impact}. \footnote{The scaling factor $\gamma_{\phi}$ models the jitter present in oscillators due to hardware imperfections \cite{sahai2013impact}}

Let $\Omega_{X}$ be the average received signal power of the signal-of-interest (SOI). The average received signal power is defined based on the free space path loss model \cite[Eq. (2.7)]{goldsmith2005wireless} and it is defined as
\begin{eqnarray} \label{Omega_x_soi}
\Omega_{X} \propto \frac{P_t}{{\left(\frac{4\cdot\pi\cdot{10^9}}{3\cdot10^8}\right)^2}{\cdot}f_{c}^2{\cdot}d_{1,g}^{2}{\cdot}\sigma_g^2}_,
\end{eqnarray}
where $P_{t}$, $d$, and $f_{c}$ are the transmit power (Watts), distance (Km), and carrier frequency (MHz), respectively. The received signal power levels are normalized with the receiver noise variance ($\sigma_g^2$). The channel between AS-1 and GS is selected as the reference link and the average received signal power in the other links are expressed relative to the reference link via the multiplicative factor $\alpha_{i,j}$,  defined as
\begin{eqnarray} \label{alpha_i_j}
\alpha_{i,j} & = & \bigg(\frac{d_{1,g}}{d_{i,j}}\bigg)^n, i\in\left\{g,1\right\}, j\in\left\{g,2\right\}, i \neq j.
\end{eqnarray}

For the case of $\alpha_{g,g}$, the variable $\alpha_{g,g}$ is treated as a scaling factor for the average residual SI power at GS. From (\ref{Omega_x_soi}) and (\ref{alpha_i_j}), the average received power of the residual SI at GS can be expressed as $\Omega_X\alpha_{g,g}$, where it is assumed that the residual SI scaled by $\alpha_{g,g}$ is below the saturation level of the FD transceiver.

From \cite{zlatanov2017capacity}, the overall level of SI suppression, i.e., combination of passive suppression with analog and digital SI cancellation, can be calculated as $\frac{1}{\alpha_{g,g}\epsilon\sigma_g^2}$.

\subsection{Air-Station 2}

Let $h_{g,2}[t]$ be the channel between GS and AS-2, $h_{1,2}[t]$ be the channel between AS-1 and AS-2, and $w_{2}[t]$ be the AWGN at AS-2 with zero mean and variance $\sigma^2_2$. From the perspective of AS-2, $x_{gs}[t]$ and $x_1[t]$ are the SOI and interfering signal, respectively. The received signal at AS-2 can be expressed as
\begin{eqnarray} \label{y_as2}
y_{2}[t] \hspace{-0.05cm} = \hspace{-0.05cm} \sqrt{\Omega_{X}\alpha_{g,2}}h_{g,2}[t]x_{gs}[t] \hspace{-0.05cm} + \hspace{-0.05cm} \sqrt{\Omega_{X}\alpha_{1,2}}h_{1,2}[t]x_{1}[t] \hspace{-0.05cm} + \hspace{-0.05cm} w_{2}[t], \hspace{-0.15cm}
\end{eqnarray}
where $\Omega_{X}\alpha_{g,2}$ and $\Omega_{X}\alpha_{1,2}$ indicate the average received signal powers of the SOI and interfering signal, respectively. 

To handle the interference at AS-2, two approaches are studied. The first approach assumes an II detector at AS-2. The II detector treats $x_1[t]$ as noise. Therefore, interference is effectively ignored. The second approach assumes a SIC detector at AS-2. The two-stage SIC detector first tries to detect and cancel $x_1[t]$ before proceeding to detect $x_{gs}[t]$ \cite{narasimhan2007individual}. 

\section{Calculation of Outage Probabilities} 
To begin the outage analysis at GS and AS-2, we first define the HBD transmission rates of AS-1 and GS as $R^{HBD}_{1}$ and $R^{HBD}_{gs}$, respectively, and the sum rate of the HBD system as $R^{HBD}_{sum} = R^{HBD}_{1}+R^{HBD}_{gs}$. Similarly, the HD transmission rates of AS-1 and GS are defined as $R^{HD}_{1}$ and $R^{HD}_{gs}$, respectively, and the sum rate of the HD system is defined as $R^{HD}_{sum} = R^{HD}_{1}+R^{HD}_{gs}$. For fair comparison between HBD and HD systems, $R_{i}^{HBD}=\frac{1}{2}R_{i}^{HD}$ for $ i \in \{1, gs\}$ \cite{kwon2010optimal,baranwal2013outage,sofotasios2017full}. The respective HBD and HD outage probabilities at GS and AS-2 are defined in the following subsections.

\subsection{Hybrid-Duplex Outage Probability}
The FD-enabled GS receives $x_1[t]$ while simultaneously transmitting $x_{gs}[t]$ in the same time slot. The simultaneous transmission and reception of signals result in strong SI at GS. Let $X_{1}=\Omega_X|h_{1,g}|^2$ be the instantaneous received signal power of the SOI at GS, modeled as a non-centered chi-squared distributed RV with Rician $K$ factor $K_{X_{1}}$. Let $Y_{si,1}=\Omega_X\alpha_{g,g}\gamma_{\phi}^2|h_{si}|^2$ and $Y_{si,2}=\Omega_X\alpha_{g,g}|\widetilde{h}_{si}|^2$ be the instantaneous received signal power corresponding to SI components. In particular, $Y_{si,1}$ is modeled as a non-centered chi-squared distributed RV with Rician $K$ factor $K_{Y_{si,1}}$ and $Y_{si,2}$ is modeled as a exponentially distributed RV.

Concurrently, AS-2 also experiences interference from AS-1. Let $X_{gs} = \Omega_{X}\alpha_{g,2}|h_{g,2}|^2$ and $Y_{1}=\Omega_{X}\alpha_{1,2}|h_{1,2}|^2$ be the instantaneous received signal power of the SOI and interference at AS-2, respectively, where $X_{gs}$ and $Y_{1}$ are independent non-centered chi-squared distributed RV with respective Rician $K$ factors $K_{X_{gs}}$ and $K_{Y_1}$, respectively. Additionally, let $\alpha\big(q,\Omega,K,\gamma\big)$ be defined as
\begin{eqnarray} \label{alpha_func}
\alpha\big(q,\Omega,K,\gamma\big) & \equiv & (-1)^q \exp(-K) \frac{{L_q}^{(0)}(K)}{(1+q)!} \Bigg(\frac{(1+K)}{\Omega}\gamma\Bigg)^{q+1}_,
\end{eqnarray}
where $q$, $\Omega$, $K$ and $\gamma$ represent an arbitrary non-negative integer, average received power of the signal, Rician $K$ factor, and threshold, respectively. The function ${L_q}^{(0)}(\bullet)$ represents the $q$-th degree, zero-order Laguerre polynomials \cite{andras2011generalized} while $\alpha\big(q,\Omega,K,\gamma\big)$ in (\ref{alpha_func}) represents the Rician power cumulative distribution function (CDF) expansion due to Rician faded signal parameters. 

\subsubsection{Ground Station}
At GS, let the HBD threshold be $\gamma_{th,gs}^{HBD} = 2^{R_{1}^{HBD}}-1$, with HBD outage event $\mathcal{O}_{gs}^{HBD} = \Big\{ h_{1,g}, h_{si} : R_{1}^{HBD} \geq \log_{2}\Big(1 + \frac{X_{1}}{Y_{si,1} + Y_{si,2} + 1}\Big)\Big\}$. By substituting $X_{1}, Y_{si,1}$ and $Y_{si,2}$ into \cite[Eq. (12)]{rached2017unified}, the closed-form outage probability at GS can be expressed as
\begin{eqnarray} \label{P_out_gs_II}
Pr\big(\mathcal{O}_{gs}^{HBD}\big) & = & \sum_{q\geq0}\sum_{l_1+l_2+l_3=q+1} \alpha\big(q,\Omega_X,K_{X_1},\gamma_{th,gs}^{HBD}\big) \nonumber\\
 & & \hspace{1cm} \times \frac{(q+1)!}{l_1! \cdot l_2! \cdot l_3!} E\{Y_{si,1}^{l_1}\} E\{Y_{si,2}^{l_2}\}_,
\end{eqnarray}
where $\alpha\big(q,\Omega_X,K_{X_1},\gamma_{th,gs}^{HBD}\big)$ is the Rician SOI power CDF expansion at GS, as defined in (\ref{alpha_func}). In addition, $E\{Y_{si,1}^{l_1}\}$ and $E\{Y_{si,2}^{l_2}\}$ are the $l_1^{th}$ and $l_2^{th}$ moments of $Y_{si,1}$ and $Y_{si,2}$, respectively. From \cite[Table II]{rached2017unified}, $E\{Y_{si,1}^{l_1}\} = \Gamma(1+l_1) \Big(\frac{\alpha_{g,g}\gamma_{\phi}^2}{1+K_{Y_{si,1}}}\Big)^{l_1}{}_1{F_1}(-l_1,1;-K_{Y_{si,1}})(\Omega_X)^{l_1}$ and $E\{Y_{si,2}^{l_2}\} = \Gamma(1+l_2) (\alpha_{g,g}\epsilon\big)^{l_2} (\Omega_X)^{l_2}$. The function ${}_1{F_1}(\bullet)$ represents the confluent Hypergeometric function \cite[Eq. (9.210.1)]{gradshteyn2014table} and summation on the right hand side (RHS) of (\ref{P_out_gs_II}) is convergent if $\gamma_{th,gs}^{HBD} \leq \frac{\Omega_X}{3(1+K_{X_{1}})(\Omega_X\alpha_{g,g}\epsilon)}$ \cite[Eq. (14)]{rached2017unified}. In (\ref{P_out_gs_II}), $E\{Y_{si,1}^{l_1}\}$ and $E\{Y_{si,2}^{l_2}\}$ quantifies the strength of residual SI due to phase noise and SI channel estimation errors, respectively. We do not expect $\alpha_{g,g}$ to approach infinity as the distance on the SI link ($d_{g,g}$) cannot be zero. However, it is possible for the average received SI power to be strong if $d_{g,g}$ is short. From $E\{Y_{si,1}^{l_1}\}$ and $E\{Y_{si,2}^{l_2}\}$, the impact of residual SI is diminished as $\alpha_{g,g} \to 0$ and hence, proper SI mitigation strategies is crucial at the FD-enabled GS. 

\subsubsection{Air-Station 2 (Interference Ignorant Detector)} \label{AS2_II_subsect}
At AS-2, let the HBD threshold be $\gamma_{th,2}^{HBD} = 2^{R_{gs}^{HBD}}-1$ and the HBD outage event be $\mathcal{O}_{2}^{HBD(II)}  = \Big\{ h_{g,2}, h_{1,2} : R_{gs}^{HBD} \geq \log_{2}\Big(1+\frac{X_{gs}}{Y_{1} + 1}\Big)\Big\}$. By substituting $X_{gs}$ and $Y_{1}$ into \cite[Eq. (12)]{rached2017unified}, the closed-form outage probability at AS-2 can be expressed as
\begin{eqnarray} \label{P_out_as2_II}
 & & \hspace{-1.5cm} Pr\big(\mathcal{O}_{2}^{HBD(II)}\big) \nonumber \\
 & = & \sum_{q\geq0} \sum_{l=0}^{q+1} \alpha\big(q,\Omega_X \alpha_{g,2}, K_{X_{gs}}, \gamma_{th,2}^{HBD}\big) \binom{q+1}{l} E\{Y_1^l\}_,
\end{eqnarray}
where $\alpha\big(q,\Omega_X \alpha_{g,2}, K_{X_{gs}}, \gamma_{th,2}^{HBD}\big)$ is the Rician SOI power CDF expansion at AS-2, as defined in (\ref{alpha_func}), and $E\{Y_1^l\}$ is the $l^{th}$ moment of the interfering signal from AS-1. From \cite[Table II]{rached2017unified}, $E\{Y_1^l\} = \Gamma(1+l) \left[\frac{\alpha_{1,2}}{1+K_{Y_1}}\right]^{l} {}_1{F_1}(-l,1;-K_{Y_1}) (\Omega_X)^{l}$ and the RHS of (\ref{P_out_as2_II}) is convergent if $\gamma_{th,2}^{HBD} \leq \frac{\Omega_{X}\alpha_{g,2}(1+K_{Y_{1}})}{2(1+K_{X_{gs}})\Omega_X\alpha_{1,2}}$ \cite[Eq. (14)]{rached2017unified}. In (\ref{P_out_as2_II}), $E\{Y_1^l\}$ quantifies the strength of the interference from AS-1 through moment parameters of $Y_1$. 

To investigate the impact of inter-AS interference, we evaluate $\lim_{\alpha_{1,2} \to L} E\{Y_1^l\}$ for $L \in \{0,\infty\}$. Although $\alpha_{1,2}$ does not reach infinity in practice, large values of $\alpha_{1,2}$ are possible when $d_{1,2}$ is small and vice-versa. Evaluating $\lim_{\alpha_{1,2} \to L} E\{Y_1^l\}$ for $L \in \{0,\infty\}$ shows that the impact of inter-AS interference reduces as $\alpha_{1,2} \to 0$, and increases as $\alpha_{1,2} \to \infty$. Thus, the II detector operates effectively in low interference scenarios such as over remote airspace where inter-AS distance is long.

\subsubsection{Air-Station 2 (Successive Interference Cancellation Detector)}

In the case of SIC, if the first stage is unable to detect the interfering signal or if the SOI cannot be detected at the second stage, then outage occurs. Therefore, the HBD outage event at AS-2 is defined as 
\begin{eqnarray}
\mathcal{O}_{2}^{HBD(SIC)} & = & \bigg\{h_{g,2}, h_{1,2} : R_{1}^{HBD} > \log_{2}\bigg(1+\frac{Y_{1}}{1+X_{gs}}\bigg) \bigg\} \nonumber\\
& \cup & \bigg\{h_{g,2}, h_{1,2} :  R_{1}^{HBD} \leq \log_{2}\bigg(1+\frac{Y_{1}}{1+X_{gs}}\bigg) \nonumber \\
 & & \hspace{1.6cm} , R_{gs}^{HBD} > \log_{2}\big(1+X_{gs}\big) \bigg\}_.
\end{eqnarray}

\begin{theorem}
The closed-form expression for outage probability with SIC detector at AS-2 is
\begin{eqnarray} \label{P_out_as2_SIC}
 & & \hspace{-1.25cm} Pr\big(\mathcal{O}_{2}^{HBD(SIC)}\big) \nonumber\\
 & = & \sum_{q\geq0}\sum_{l=0}^{q+1} \alpha\big(q,\Omega_X\alpha_{1,2},K_{Y_{1}},\gamma_{th,gs}^{HBD}\big) \binom{q+1}{l} E\{X_{gs}^l\} \nonumber \\
 & & + 1 - Q_1\Bigg( \sqrt{2K_{X_{gs}}}, \sqrt{\frac{2(K_{X_{gs}}+1)\gamma_{th,2}^{HBD}}{\Omega_{X}\alpha_{g,2}}} \Bigg) \nonumber\\
 & & - \sum_{n\geq0}\sum_{i=0}^{n}\sum_{j=0}^{i+1} \alpha\big(i,\Omega_X\alpha_{1,2},K_{Y_{1}},\gamma_{th,gs}^{HBD}\big) \nonumber\\
 & & \times \alpha\big(n-i,\Omega_X\alpha_{g,2},K_{X_{gs}},1\big)  \binom{i+1}{j}\frac{(\gamma_{th,2}^{HBD})^{j+n-i+1}}{j+n-i+1}_,
\end{eqnarray}
where $Q_1\left(\cdot,\cdot\right)$ is the Marcum Q function \cite{andras2011generalized}, \cite[Eq. (4.33)]{simon2005digital} and $E\{X_{gs}^l\}=\Gamma(1+l)\left(\frac{\Omega_{X}\alpha_{g,2}}{1+K_{X_{gs}}}\right)^{l} {}_1{F_1}(-l,1;-K_{X_{gs}})$ \cite[Table II]{rached2017unified}.
\end{theorem}

\begin{IEEEproof}
The proof can be found in Appendix \ref{SIC_proof}.
\end{IEEEproof}
The first term in (\ref{P_out_as2_SIC}) is the outage probability due to detecting interference from AS-1. The second term in (\ref{P_out_as2_SIC}) is the outage probability due to SOI detection after interference cancellation. From $\alpha\big(q,\Omega_X\alpha_{1,2},K_{Y_{1}},\gamma_{th,2}^{HBD}\big)$ in (\ref{P_out_as2_SIC}), it is evident that the SIC detector works effectively in high interference scenarios, such as in congested airspace where inter-AS distance is short, since the effect of interference at the SOI detection stage is diminished when $\alpha_{1,2}$ is large. The closed-form expressions in (\ref{P_out_gs_II}), (\ref{P_out_as2_II}) and (\ref{P_out_as2_SIC}) can shed insights into the impact of residual SI at GS and interference from AS-1 at AS-2. Further discussions on outage performance with respect to the level of interference are presented in Section V. 

\subsection{Half-Duplex Outage Probability}
When the GS is operating in HD mode, AS-2 does not experience interference from AS-1. Let the HD threshold at GS and AS-2 be defined as $\gamma_{th,gs}^{HD} = 2^{2R_{1}^{HBD}}-1$ and $\gamma_{th,2}^{HD} = 2^{2R_{gs}^{HBD}}-1$, respectively. Then, the HD outage probabilities at GS and AS-2 are given in (\ref{P_out_hd_gs}) and (\ref{P_out_hd_as2}), respectively \cite[Table I]{rached2017unified}.
\begin{eqnarray} 
Pr\big(\mathcal{O}_{gs}^{HD}\big) & = & \sum_{m\geq0} \alpha\big(m,\Omega_X,K_{X_1},\gamma_{th,gs}^{HD}\big)_, \label{P_out_hd_gs} \\
Pr\big(\mathcal{O}_{2}^{HD}\big) & = & \sum_{m\geq0} \alpha\big(m,\Omega_X\alpha_{g,2},K_{X_{gs}},\gamma_{th,2}^{HD}\big)_. \label{P_out_hd_as2}
\end{eqnarray}

The outage probability expressions in (\ref{P_out_hd_gs}) and (\ref{P_out_hd_as2}) can be used as a benchmark comparison against HBD mode at GS and AS-2, respectively, which is presented in Section V. 

\subsection{System Level Outage Probability}
For the proposed multi-user system, the overall system level outage probability is used as a performance metric to compare HBD and HD protocols. For $\beta \in \{HBD(II), HBD(SIC)\}$, the system level outage probability is defined as $P_{out,system}^{\beta} = \max\left(Pr\big(\mathcal{O}_{gs}^{HBD}\big),Pr\big(\mathcal{O}_{2}^{\beta}\big)\right)$ and $P_{out,system}^{HD} = \max\left(Pr\big(\mathcal{O}_{gs}^{HD}\big),Pr\big(\mathcal{O}_{2}^{HD}\big)\right)$. The system level outage probability provides the worst case system level outage behavior for the II and SIC detectors and allows the identification of performance bottlenecks in HBD-ACS. Having knowledge of the performance bottleneck in the HBD-UCS enables interference management to be more effective. For instance, if the link between the FD-enables GS and AS-2 has the highest outage probability, then the design requirements of the SI mitigation architecture at the FD-enabled GS can be less stringent which can lead to hardware with lower cost or power requirements.


\section{Finite SNR Analysis}

In the following subsections, the mathematical preliminaries and derivations related to finite SNR diversity gain are presented for both fixed and variable transmission rates, with detailed derivation omitted for brevity. As it will be shown, finite SNR analysis is an effective tool in evaluating the performance of the II and SIC detectors in an interference-limited environment.
  
\subsection{Mathematical Preliminaries}
\subsubsection{Finite SNR Diversity Gain}

For a system with outage event $\mathcal{O}$, outage probability $Pr\big(\mathcal{O}\big)$, transmission rate $R$, threshold $\gamma$, and average received power $\Omega$ with unit noise variance, the diversity gain $d$ at high SNR is given by Zheng and Tse \cite{zheng2003diversity} as
\begin{eqnarray} \label{asymp_diversity_gain}
 d = \lim_{\Omega\to\infty} \frac{\log_2(Pr\big(\mathcal{O}\big))}{\log_2(\Omega)}_.
\end{eqnarray}

The diversity gain definition in (\ref{asymp_diversity_gain}) is for systems that operate at high SNR ranges. The finite SNR diversity gain $d_f$, which quantifies the decay rate of the outage probability at low-to-moderate SNRs, is given as \cite[Eq. (5)]{narasimhan2006finite}
\begin{eqnarray} \label{df}
d_f = \frac{-\Omega}{Pr\big(\mathcal{O}\big)}\frac{\partial}{\partial\Omega}Pr\big(\mathcal{O}\big)_.
\end{eqnarray}

It has since been shown by Shin et al. \cite{shin2008diversity} and Heidarpour et al. \cite{heidarpour2017finite} that $\lim_{\Omega\to\infty}d_f = d$. Therefore, (\ref{df}) is consistent with the asymptotic diversity definitions in \cite{zheng2003diversity} at high SNR. Practical wireless systems typically operate at the low-to-moderate SNR range \cite{narasimhan2006finite}. The outage behavior of these systems may also be different at high and moderate SNRs. Therefore, there is motivation to quantify diversity gains at finite SNRs since (\ref{asymp_diversity_gain}) does not accurately reflect outage behaviors at low-to-moderate SNRs \cite{shin2008diversity}. 

\subsubsection{Finite SNR DMT Parameters}
For a system which varies its transmission rate with respect to $\Omega$, i.e., variable transmission rate, the high SNR multiplexing gain $r$ is given by Zheng and Tse \cite{zheng2003diversity} as
\begin{eqnarray} \label{asymp_mult_gain}
r = \lim_{\Omega\to\infty} \frac{R(\Omega)}{\log_2(\Omega)}
\end{eqnarray}
and the finite SNR multiplexing gain $r_f$ for such systems is \cite[Eq. (4)]{narasimhan2006finite}
\begin{eqnarray} \label{rf}
r_f = \frac{R(\Omega)}{\log_2(1+\Omega)}_.
\end{eqnarray}

It has similarly been shown by Shin et al. \cite{shin2008diversity} and Heidarpour et al. \cite{heidarpour2017finite} that $\lim_{\Omega\to\infty}r_f = r$, with $Pr\big(\mathcal{O}\big)$ computed with respect to the threshold $\gamma=(1+\Omega)^{r_f}-1$. The finite SNR diversity gain for such a variable transmission rate system (denoted as $d_{f}^{*}$) can be obtained from (\ref{df}) as \cite[Eq. (36)]{shin2008diversity}
\begin{eqnarray}
d_f^* & = & \frac{-\Omega}{Pr\big(\mathcal{O}\big)} \lim_{\Delta(\Omega)\to0} \bigg[\frac{Pr\big(\mathcal{\widehat{O}}\big) - Pr\big(\mathcal{O}\big)}{\Delta(\Omega)}\bigg]_, \label{df_var_rate1} 
\end{eqnarray}
where $\Delta(\Omega)=\widehat{\Omega}-\Omega$, $\widehat{\Omega}>\Omega$ and $\mathcal{\widehat{O}}$ is the outage event with respect to $R\big(\Omega+\Delta(\Omega)\big)$. Furthermore, $Pr\big(\mathcal{\widehat{O}}\big)$ is the outage probability with average received power $\widehat{\Omega}=\Omega+\Delta(\Omega)$, threshold $\widehat{\gamma}=[1+\Omega+\Delta(\Omega)]^{r_f}-1$ and $Pr\big(\mathcal{\widehat{O}}\big)=Pr\big(\mathcal{O}\big)$ when $\Delta(\Omega)=0$. Applying L'Hospital's rule in (\ref{df_var_rate1}) by differentiating with respect to $\Delta(\Omega)$ and setting $\Delta(\Omega)=0$ yields
\begin{eqnarray} 
d_f^* & = & \frac{-\Omega}{Pr\big(\mathcal{O}\big)} \frac{\partial}{\partial\Delta(\Omega)} Pr\big(\mathcal{\widehat{O}}\big) {\Bigg|_{\Delta(\Omega)=0}}_. \label{df_var_rate2}
\end{eqnarray}

Let $Z$ be a RV with normalized $n^{th}$ moment defined as $M\{Z^{n}\}=\frac{E\{Z^{n}\}}{(\Omega)^{n}}$ and let function $g(i,j,\Omega,r_f)$ be 
\begin{eqnarray} 
g(i,j,\Omega,r_f) & = & \big(\big[1 + \Omega\big]^{r_f}-1\big)^{i} (\Omega)^{j} \nonumber \\
 & & \hspace{-2cm} \times \bigg[ \left(\big[1 + \Omega\big]^{r_f} - 1\right) \frac{j}{\Omega} + (i+1)(r_f)(1 + \Omega)^{r_f-1} \bigg]_,
\end{eqnarray}
where $i$ and $j$ are integers. The function $M\{Z^{n}\}$ represents the normalized $n^{th}$ moment of an interfering signal while the function $g(i,j,\Omega,r_f)$ reflects the outage probability decay rate of a variable transmission rate scheme due to average received power ($\Omega$) and finite SNR multiplexing gain ($r_f$).

Although \cite[Eq. (36)]{shin2008diversity} and \cite[Eq. (5)]{narasimhan2006finite} evaluate finite SNR diversity gains using different approaches, the principles underlying them are the same since the latter is an extension of the former. To this end, (\ref{df_var_rate2}) can be used to evaluate $d_f^*$ for adaptive systems, with $r_f$ indicating the sensitivity of the rate adaptation scheme \cite{narasimhan2006finite}. It is also of interest to analyze $d_f^*$ as it can lead to better code designs that improve transmission rates at the expense of reliability for adaptive systems and vice-versa.

\subsection{Finite SNR Diversity Gain for HBD Systems}
Let the finite SNR HBD diversity gain at GS and AS-2 be defined as $d_{f,gs}^{HBD}$ and $d_{f,2}^{HBD,i}, i \in \{II,SIC\}$, respectively. Additionally, let $R_1$ and $R_{gs}$ be fixed constants with average received power $\Omega = \Omega_X$. Then, the finite SNR diversity gain at GS and AS-2 are presented in the following propositions.

\begin{proposition}
The finite SNR diversity gain at the FD-enabled GS is
\begin{eqnarray} \label{df_fixed_GS}
d_{f,gs}^{HBD} & = & \frac{-\Omega_X}{Pr\big(\mathcal{O}_{gs}^{HBD}\big)} \sum_{q\geq0}\sum_{l_1+l_2+l_3=q+1} \alpha\big(q,1,K_{X_1},\gamma_{th,gs}^{HBD}\big) \nonumber \\
 & & \hspace{-1.9cm} \times \frac{(q+1)!(l_1+l_2-q-1)}{l_1! \cdot l_2! \cdot l_3!} M\{Y_{si,1}^{l_1}\} M\{Y_{si,2}^{l_2}\}  \left(\Omega_X\right)^{l_1+l_2-q-2}_.
\end{eqnarray}
\end{proposition}
\begin{IEEEproof}
The finite SNR diversity gain at GS can be obtained by substituting (\ref{P_out_gs_II}) into (\ref{df}).
\end{IEEEproof}

At low-to-moderate $\Omega_X$, the outage behavior at GS can be analyzed from (\ref{df_fixed_GS}). In particular, (\ref{df_fixed_GS}) allows observation of subtle changes in outage behavior due to the scaling factor associated with the SI strength ($\alpha_{g,g}$) and SI channel estimation error ($\epsilon$) that is not present at high $\Omega_X$. In addition, the asymptotic behavior of $d_{f,gs}^{HBD}$ can be obtained from (\ref{df_fixed_GS}) as shown in the following corollary.

\begin{corollary} \label{coro_asymp_df_fixed_GS}
The asymptotic behavior of $d_{f,gs}^{HBD}$ is given by
\begin{eqnarray} \label{asymp_df_fixed_GS}
\lim_{\Omega_X\to\infty} \frac{-\Omega_X}{Pr\big(\mathcal{O}_{gs}^{HBD}\big)}\frac{\partial}{\partial\Omega_X}Pr\big(\mathcal{O}_{gs}^{HBD}\big) & = & 0.
\end{eqnarray}
\begin{IEEEproof}
The proof is given in Appendix \ref{coro_asymp_df_fixed_GS_proof}.
\end{IEEEproof}
\end{corollary}

From (\ref{asymp_df_fixed_GS}), $d_{f,gs}^{HBD} \to 0$ as $\Omega_X \to \infty$ because increasing $\Omega_X$ also causes residual SI to be stronger, hence there is no improvement in the overall SINR. Also, (\ref{asymp_df_fixed_GS}) suggests that the tolerance for residual SI in HBD-ACS is progressively diminished as $\Omega_X$ is increased since $d_{f,gs}^{HBD} \to 0$ corresponds to negligible improvements in outage probability at GS.

\begin{proposition}
The finite SNR diversity gain at AS-2 with the II ($d_{f,2}^{HBD(II)}$) and SIC detectors ($d_{f,2}^{HBD(SIC)}$) are
\begin{eqnarray} 
d_{f,2}^{HBD(II)} & \hspace{-0.2cm} = & \hspace{-0.2cm} \frac{-\Omega_X}{Pr\big(\mathcal{O}_{2}^{HBD(II)}\big)} \sum_{q\geq0} \sum_{l=0}^{q+1} \alpha\big(q,\alpha_{g,2}, K_{X_{gs}}, \gamma_{th,2}^{HBD}\big) \nonumber \\
 & & \times \binom{q+1}{l} M\{Y_1^l\} (l-q-1) (\Omega_X)^{l-q-2}_, \label{df_fixed_AS2_II} \\
d_{f,2}^{HBD(SIC)} & \hspace{-0.2cm} = & \hspace{-0.2cm} \frac{-\Omega_X}{Pr\big(\mathcal{O}_{2}^{HBD(SIC)}\big)} \Bigg[  \sum_{q\geq0}\sum_{l=0}^{q+1} \alpha\big(q,\alpha_{1,2},K_{Y_{1}},\gamma_{th,gs}^{HBD}\big) \nonumber \\
 & & \times \binom{q+1}{l} M\{X_{gs}^l\} (l-q-1) (\Omega_{X})^{l-q-2} \nonumber\\
 & & \hspace{-2cm} + \sum_{m\geq0} \alpha\big(m,\alpha_{g,2},K_{X_{gs}},\gamma_{th,2}^{HBD}\big) (-m-1) (\Omega_{X})^{-m-2} \nonumber \\
 & & \hspace{-2cm} - \sum_{n\geq0}\sum_{i=0}^{n}\sum_{j=0}^{i+1} \alpha\big(i,\alpha_{1,2},K_{Y_{1}},\gamma_{th,gs}^{HBD}\big) \alpha\big(n-i,\alpha_{g,2},K_{X_{gs}},1\big) \nonumber\\
 & & \hspace{-0.3cm} \times \binom{i+1}{j} \frac{\big(\gamma_{th,2}^{HBD}\big)^{j+n-i+1}}{j+n-i+1} (-n-2) (\Omega_X)^{-n-3} \Bigg]_. \label{df_fixed_AS2_SIC}
\end{eqnarray}
\end{proposition}
\begin{IEEEproof}
At AS-2, $d_{f,2}^{HBD(i)}, i \in \{II,SIC\}$ can be obtained for the II and SIC by respectively substituting (\ref{P_out_as2_II}) and (\ref{P_out_as2_SIC}) into (\ref{df}).
\end{IEEEproof}

The outage behavior at AS-2 can be analyzed from (\ref{df_fixed_AS2_II}) and (\ref{df_fixed_AS2_SIC}) at low-to-moderate $\Omega_X$. In particular, (\ref{df_fixed_AS2_II}) and (\ref{df_fixed_AS2_SIC}) enables the observation of subtle changes in outage behavior for both II and SIC detectors, which are not present at high $\Omega_X$, as inter-aircraft interference varies. Extending upon (\ref{df_fixed_AS2_II}) and (\ref{df_fixed_AS2_SIC}),  the asymptotic behavior of $d_{f,2}^{HBD(i)}, i \in \{II,SIC\}$ can be obtained as follows.

\begin{corollary} \label{coro_asymp_df_fixed_AS2}
The asymptotic behavior of $d_{f,2}^{HBD(i)}, i \in \{II,SIC\}$ is given by
\begin{eqnarray} \label{asymp_df_fixed_AS2}
\lim_{\Omega_X\to\infty} \frac{-\Omega_X}{Pr\big(\mathcal{O}_{2}^{HBD,i}\big)}\frac{\partial}{\partial\Omega_X}Pr\big(\mathcal{O}_{2}^{HBD,i}\big) & = & 0.
\end{eqnarray}
\begin{IEEEproof}
The proof is given in Appendix \ref{coro_asymp_df_fixed_AS2_proof}.
\end{IEEEproof}
\end{corollary}

\begin{remark}
\textit{In \cite{etkin2008gaussian} and \cite{sirigina2016symbol}, the average received signal powers of the desired ($\Omega_x \alpha_{g,2}$) and interfering ($\Omega_x \alpha_{1,2}$) links are related through an exponent, where a large exponent corresponds to very strong interference. At high SNRs, the SIC-based receiver is shown to achieve full diversity under very strong interference levels. In contrast, this work demonstrates that the SIC detector achieves zero diversity gain at high SNRs when the desired and interfering signal levels are related through a scaling parameter.}
\end{remark}

In the presence of interference at AS-2, (\ref{asymp_df_fixed_AS2}) shows that improvements to outage probability at AS-2 progressively diminishes since $d_{f,2}^{HBD(i)} \to 0$ as $\Omega_X \to \infty$ for $i \in \{II,SIC\}$. For the II detector, increasing $\Omega_X$ results in strong interference. As a consequence, there is no improvement to the overall SINR. Hence, the II detector is unsuitable in strong interference environments. Similarly, for the SIC detector, increasing $\Omega_X$ causes $x_{gs}[t]$ to be stronger, making the detection and subtraction of $x_1[t]$ increasingly challenging at stage 1 of the SIC detector. Hence, $\alpha_{1,2}$ must either increase (for the II detector) or decrease (for the SIC detector) at high $\Omega_X$ for HBD-ACS to see meaningful improvements in outage probability.

\subsection{Finite SNR Diversity Gain for HD Systems}

Let the finite SNR diversity gain at GS and AS-2 be defined as $d_{f,i}^{HD}, i \in \{gs,2\}$, respectively, with $R_1$ and $R_{gs}$ assumed to be constants with average received power $\Omega = \Omega_X$. Then, the finite SNR diversity gain at GS and AS-2 are presented in the following proposition.

\begin{proposition}
The finite SNR diversity gain at GS and AS-2 operating in HD mode are given in (\ref{df_fixed_hd_GS}) and (\ref{df_fixed_hd_as2}), respectively.
\begin{eqnarray} 
d_{f,gs}^{HD} & = & \frac{-\Omega_X}{Pr\big(\mathcal{O}_{gs}^{HD}\big)} \sum_{m\geq0} \alpha\big(m,1,K_{X_1},\gamma_{th,gs}^{HD}\big) \nonumber \\
 & & \hspace{2.7cm} \times (-m-1) (\Omega_{X})^{-m-2}_, \label{df_fixed_hd_GS}\\
d_{f,2}^{HD} & = & \frac{-\Omega_X}{Pr\big(\mathcal{O}_{2}^{HD}\big)} \sum_{m\geq0} \alpha\big(m,\alpha_{g,2},K_{X_{gs}},\gamma_{th,2}^{HD}\big) \nonumber \\
 & &  \hspace{2.7cm} \times (-m-1) (\Omega_{X})^{-m-2}_. \label{df_fixed_hd_as2}
\end{eqnarray}
\end{proposition}
\begin{IEEEproof}
The expressions in (\ref{df_fixed_hd_GS}) and (\ref{df_fixed_hd_as2}) can be obtained by respectively substituting (\ref{P_out_hd_gs}) and (\ref{P_out_hd_as2}) into (\ref{df}).
\end{IEEEproof}

The HD outage behavior at GS and AS-2 can be analyzed from (\ref{df_fixed_hd_GS}) and (\ref{df_fixed_hd_as2}), respectively, and it enables the observation of changes in outage probability decay rate that is not visible at high $\Omega_X$. As $\Omega_X \to \infty$, (\ref{df_fixed_hd_GS}) and (\ref{df_fixed_hd_as2}) can be evaluated to determine the asymptotic diversity gain as follows.

\begin{corollary} \label{coro_lim_df_fixed_hd}
The asymptotic behavior of $d_{f,i}^{HD}, i \in \{gs,2\}$ is
\begin{eqnarray}  \label{lim_df_fixed_hd}
\lim_{\Omega_X \to \infty} d_{f,i}^{HD} & = & 1. 
\end{eqnarray}

\begin{IEEEproof}
The proof is provided in Appendix \ref{coro_lim_df_fixed_hd_proof}.
\end{IEEEproof}
\end{corollary}

From (\ref{lim_df_fixed_hd_GS}), $d_{f,i}^{HD} \to 1$ as $\Omega_X \to \infty$ for $i \in \{gs,2\}$ and it indicates that the HD system achieves full diversity in the absence of interference at high $\Omega_X$, which is consistent with \cite[Fig. 3]{shin2008diversity}.

\subsection{Finite SNR DMT Analysis for HBD Systems}

Let the finite SNR diversity gain at GS for a HBD system be defined as $d_{f,gs}^{HBD*}$, with variable transmission rate $R_1^{HBD}(\Omega_X)=r_f\log_2(1+\Omega_X)$ and threshold $\gamma_{th,gs}^{HBD} = [1+\Omega_X]^{r_f}-1$. Similarly, let the finite SNR diversity gain at AS-2 for a HBD system be denoted as $d_{f,2}^{HBD(i)*}, i \in \{II,SIC\}$, with variable transmission rate $R_{gs}^{HBD}(\Omega_X)=r_f\log_2(1+\Omega_X)$ and threshold $\gamma_{th,2}^{HBD} = [1+\Omega_X]^{r_f}-1$. The finite SNR diversity gains at GS and AS-2 are presented in the following propositions.

\begin{proposition}
At GS, the finite SNR diversity gain is given as
\begin{eqnarray} \label{df_var_gs}
d_{f,gs}^{HBD*} & = & \frac{-\Omega_X}{Pr\big(\mathcal{O}_{gs}^{HBD}\big)} \sum_{q\geq0}\sum_{l_1+l_2+l_3=q+1} \alpha\big(q,1,K_{X_1},1\big) \nonumber \\
 & & \hspace{-2.4cm} \times \frac{(q+1)!}{l_1! \cdot l_2! \cdot l_3!} M\{Y_{si,1}^{l_1}\} M\{Y_{si,2}^{l_2}\} g(q,l_1+l_2-q-1,\Omega_X,r_f).
\end{eqnarray}
\end{proposition}
\begin{IEEEproof}
Let $\Omega=\Omega_X$, $\widehat{\Omega}=\widehat{\Omega}_X$ and $\gamma = \gamma_{th,gs}^{HBD}=[1+\Omega_X]^{r_f}-1$ and $\mathcal{O} = \mathcal{O}_{gs}^{HBD}$. Then, $d_{f,gs}^{HBD*}$ can be obtained through algebraic manipulations by substituting (\ref{P_out_gs_II}) into (\ref{df_var_rate2}).
\end{IEEEproof}

\begin{proposition}
At AS-2, the finite SNR diversity gain with II and SIC detectors are
\begin{eqnarray} 
d_{f,2}^{HBD(II)*} & = & \frac{-\Omega_X}{Pr\big(\mathcal{O}_{2}^{HBD(II)}\big)} \sum_{q\geq0} \sum_{l=0}^{q+1} \alpha\big(q,\alpha_{g,2}, K_{X_{gs}}, 1\big) \nonumber \\
 & & \times \binom{q+1}{l} M\{Y_1^l\} g(q,l-q-1,\Omega_X,r_f)_, \label{df_var_AS2_II} \\
d_{f,2}^{HBD(SIC)*} & = & \frac{-\Omega_X}{Pr\big(\mathcal{O}_{2}^{HBD(SIC)}\big)} \Bigg[ \sum_{q\geq0}\sum_{l=0}^{q+1} \alpha\big(q,\alpha_{1,2},K_{Y_{1}},1\big) \nonumber \\
 & & \times \binom{q+1}{l} M\{X_{gs}^l\} g(q,l-q-1,\Omega_X,r_f) \nonumber\\
 & & \hspace{-2cm} + \sum_{m\geq0} \alpha\big(m,\alpha_{g,2},K_{X_{gs}},1\big) g(m,-m-1,\Omega_X,r_f) \nonumber \\
 & & \hspace{-2cm} - \sum_{n\geq0}\sum_{i=0}^{n}\sum_{j=0}^{i+1} \alpha\big(i,\alpha_{1,2},K_{Y_{1}},1\big) \alpha\big(n-i,\alpha_{g,2},K_{X_{gs}},1\big) \nonumber\\
 & & \hspace{-0.6cm} \times \frac{\binom{i+1}{j}}{j+n-i+1} g(j+n+1,-n-2,\Omega_X,r_f) \Bigg]_. \label{df_var_AS2_SIC}
\end{eqnarray}
\end{proposition}
\begin{IEEEproof}
Let $\Omega=\Omega_X$, $\widehat{\Omega}=\widehat{\Omega}_X$ and $\gamma=\gamma_{th,2}^{HBD}=[1+\Omega_X]^{r_f}-1$ and $\mathcal{O} = \mathcal{O}_{2}^{HBD,i}, i \in \{II,SIC\}$. Then, similar to (\ref{df_var_gs}), $d_{f,2}^{HBD(II)*}$ and $d_{f,2}^{HBD(SIC)*}$ can be obtained through algebraic manipulations by respectively substituting (\ref{P_out_as2_II}) and (\ref{P_out_as2_SIC}) into (\ref{df_var_rate2}). 
\end{IEEEproof}

In the presence of interference at GS and AS-2, DMT at low-to-moderate $\Omega_X$ can be analyzed from (\ref{df_var_gs}), (\ref{df_var_AS2_II}) and (\ref{df_var_AS2_SIC}). It reveals the interference scenarios in which the II or SIC detectors achieves better diversity gain than HD systems.

\subsection{Finite SNR DMT Analysis for HD Systems}
Let the finite SNR HD diversity gain at GS be defined as $d_{f,gs}^{HD*}$. To ensure fair comparison, we let the variable HD date rate be twice the variable HBD data rate. Let $R_1^{HD}(\Omega_X)=2r_f\log_2(1+\Omega_X)$ be the variable transmission rate at AS-1 with threshold $\gamma_{th,gs}^{HD} = [1+\Omega_X]^{2r_f}-1$. Similarly at AS-2, let the finite SNR HD diversity gain at AS-2 be defined as $d_{f,2}^{HD*}$ with variable transmission rate $R_{gs}^{HD}(\Omega_X)=2r_f\log_2(1+\Omega_X)$ and threshold $\gamma_{th,2}^{HD} = [1+\Omega_X]^{2r_f}-1$. The closed-form expressions for the finite SNR diversity gains at GS and AS-2 are presented in the following proposition.

\begin{proposition}
For a variable transmission rate scheme, the finite SNR diversity gain at GS and AS-2 are given in (\ref{df_var_gs_HD}) and (\ref{df_var_AS2_HD}), respectively.
\begin{eqnarray} 
d_{f,gs}^{HD*} & = & \frac{-\Omega_X}{Pr\big(\mathcal{O}_{gs}^{HD}\big)} \sum_{m\geq0} \alpha\big(m,1,K_{X_1},1\big) \nonumber \\
 & & \hspace{2cm} \times g(m,-m-1,\Omega_X,2r_f)_, \label{df_var_gs_HD} \\
d_{f,2}^{HD*} & = & \frac{-\Omega_X}{Pr\big(\mathcal{O}_{2}^{HD}\big)} \sum_{m\geq0} \alpha\big(m,\alpha_{g,2},K_{X_{gs}},1\big) \nonumber \\
 & & \hspace{2cm} \times g(m,-m-1,\Omega_X,2r_f)_. \label{df_var_AS2_HD}
\end{eqnarray}
\end{proposition}
\begin{IEEEproof}
The expressions in (\ref{df_var_gs_HD}) and (\ref{df_var_AS2_HD}) can be obtained through algebraic manipulations by respectively substituting (\ref{P_out_hd_gs}) and (\ref{P_out_hd_as2}) into (\ref{df_var_rate2}).
\end{IEEEproof}

In the absence of interference, (\ref{df_var_gs_HD}) and (\ref{df_var_AS2_HD}) can be used to evaluate the DMT at GS and AS-2, providing a benchmark that can be used in evaluating the performance of the II and SIC detectors in HBD systems.

\subsection{System Level Finite SNR Diversity Gain and DMT}

The system level finite SNR diversity gain and DMT for the multi-user system in Fig. \ref{fig:1} will be used as a metric to compare HBD and HD systems. For fixed transmission rate schemes, the HBD and HD system level finite SNR diversity gain are defined as $d_{f,system}^{\beta} = \min\left( d_{f,gs}^{HBD}, d_{f,2}^{\beta} \right)$ and $d_{f,system}^{HD} = \min\left( d_{f,gs}^{HD}, d_{f,2}^{HD} \right)$, respectively. Similarly, for variable transmission rate schemes, the HBD and HD system level finite SNR DMT are defined as $d_{f,system}^{\beta*} = \min\left( d_{f,gs}^{HBD*}, d_{f,2}^{\beta*} \right)$ and $d_{f,system}^{HD} = \min\left( d_{f,gs}^{HD*}, d_{f,2}^{HD*} \right)$, respectively. Quantifying the finite SNR diversity gain and DMT provides insights into the degree of improvements in outage performance at the system level, which will be further discussed in Section V.  
\section{Numerical Results}

In this section, numerical results pertaining to the outage probabilities and finite SNR diversity gains at GS, AS-2 and system level are discussed. Monte Carlo simulations are conducted with $10^{9}$ samples to verify the accuracy of the outage probability computations. In addition, all Rician $K$ factors are fixed at 15, i.e., $K_{X_1}=K_{Y_{si,1}}=K_{X_{gs}}=K_{Y_1}=15$ \cite{matolak2017air_suburban}, with $\sigma_g^2 = \sigma_2^2=-115$ dBm \footnote{Assuming a noise figure of $6$ dB, $\sigma_g^2 = \sigma_2^2=-115$ dBm results in an effective bandwidth of 200kHz. Such a bandwidth falls within the range of existing VHF datalink (VDL) standards (25kHz) \cite[Table 3.16]{stacey2008aeronautical}, and the upcoming orthogonal frequency-division multiplexing (OFDM)-based L-band digital aeronautical communication systems-1 (LDACS-1) standard \cite{jamal2017fbmc,gligorevic2011ldacs1} and Gaussian minimum shift keying (GMSK)-based LDACS-2 standard (500kHz) \cite{jamal2017fbmc}.} \cite{itu2011m2233}, $R_{sum}^{HD}=R_{sum}^{HBD}=1$ for fair comparison between the HBD and HD systems. Furthermore, a phase noise strength of $\gamma_{\phi}^2=-130$ dBm is chosen and is subsequently normalized by $\sigma_g^2 = -115$ dBm in the simulations. \footnote{It is worth noting that the normalized phase noise strength falls within the range of phase noise seen in Appendix C of \cite{sahai2013impact}.} At the FD-enabled GS, SI suppression levels of $163$ dB to $175$ dB are considered by choosing $\alpha_{g,g}=\{1, 1.5\}$ and $\epsilon = \{0.01, 0.001\}$. Finally, the subsequent analysis in this section is conducted for $0\text{ dB} \leq \Omega_X \leq 30\text{ dB}$. \footnote{In this work, we consider $0\text{ dBm} \leq P_t \leq 36.4\text{ dBm}$. Taking the GMSK-based LDACS-2 as an example \cite{jamal2017fbmc}, with 200kHz of bandwidth, noise figure of 6 dB, and carrier frequency $f_c = 968$MHz, a transmit power of $P_t = 26.4$ dBm is obtained for $d_{1,g}=9.2$km \cite{matolak2017air_suburban}, $\alpha_{1,2}=0.5$, $d_{1,2}=4.6$km \cite{caas2014manual}, and $\Omega_X = 30$ dB. When $d_{1,g}=29$km \cite{matolak2017air_suburban}, $\alpha_{1,2}=5$, $d_{1,2}=145$km \cite{caas2014manual}, and $\Omega_X = 30$ dB, $P_t = 36.4$ dBm is obtained. The obtained values of $P_t$ is comparable to \cite{jamal2017fbmc}, where a transmit power of $P_t=41$ dBm was used in the performance analysis of LDACS-2.}

\subsection{Finite SNR Diversity Gain and Outage Analysis}

\subsubsection{\underline{\textit{Impact of Residual SI at GS}}}

\begin{observation}
\emph{The FD-enabled GS has near-ideal outage probability and diversity gain at very low SNR, and is interference-limited at high SNR.}
\end{observation}

\begin{figure*}[]
\centering
\subfloat[Outage probability comparison at GS]{\includegraphics [width=0.9\columnwidth]{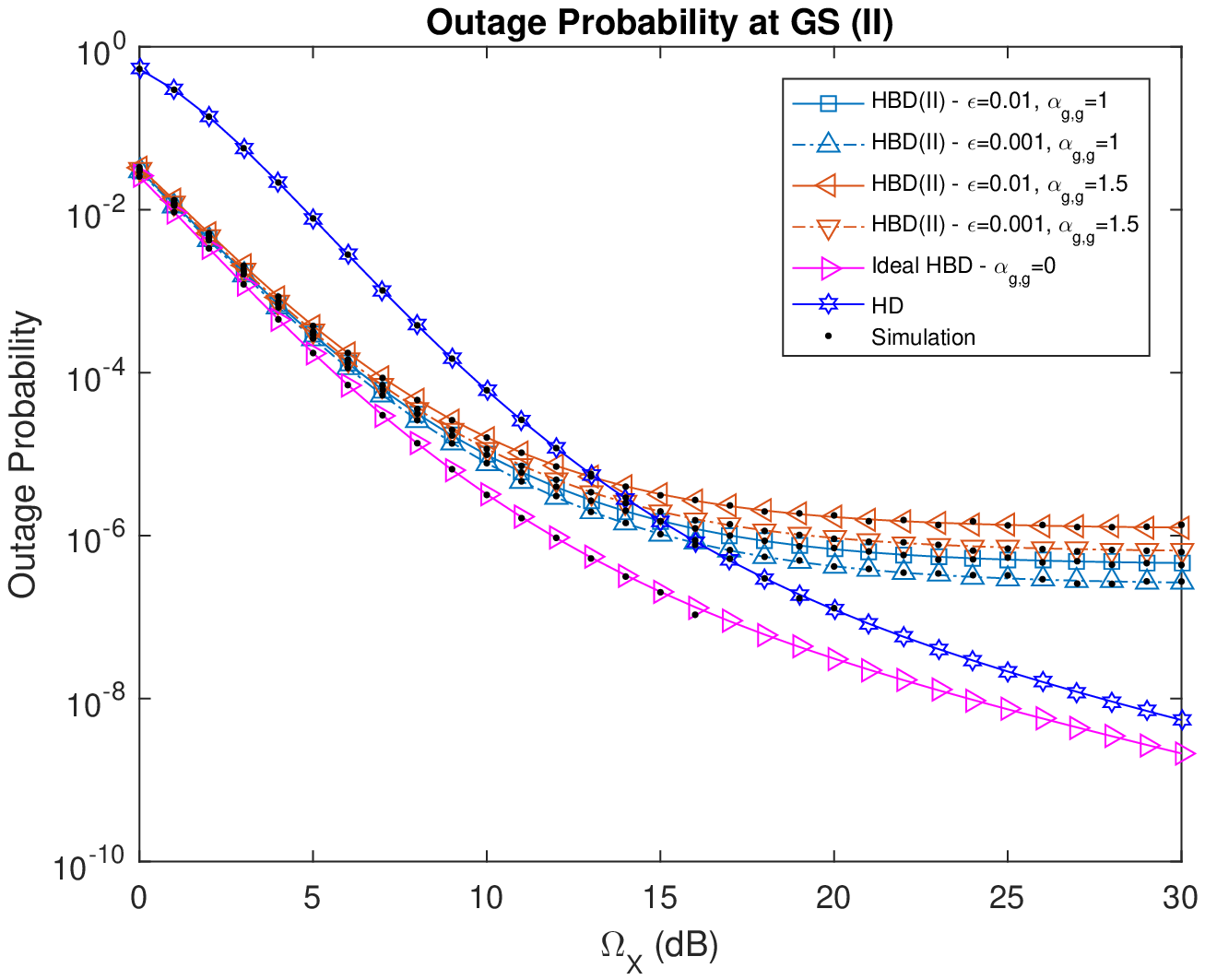}
\label{fig:fixed_pout_gs}}
\hfil
\subfloat[Finite SNR diversity gain comparison at GS]{\includegraphics [width=0.9\columnwidth]{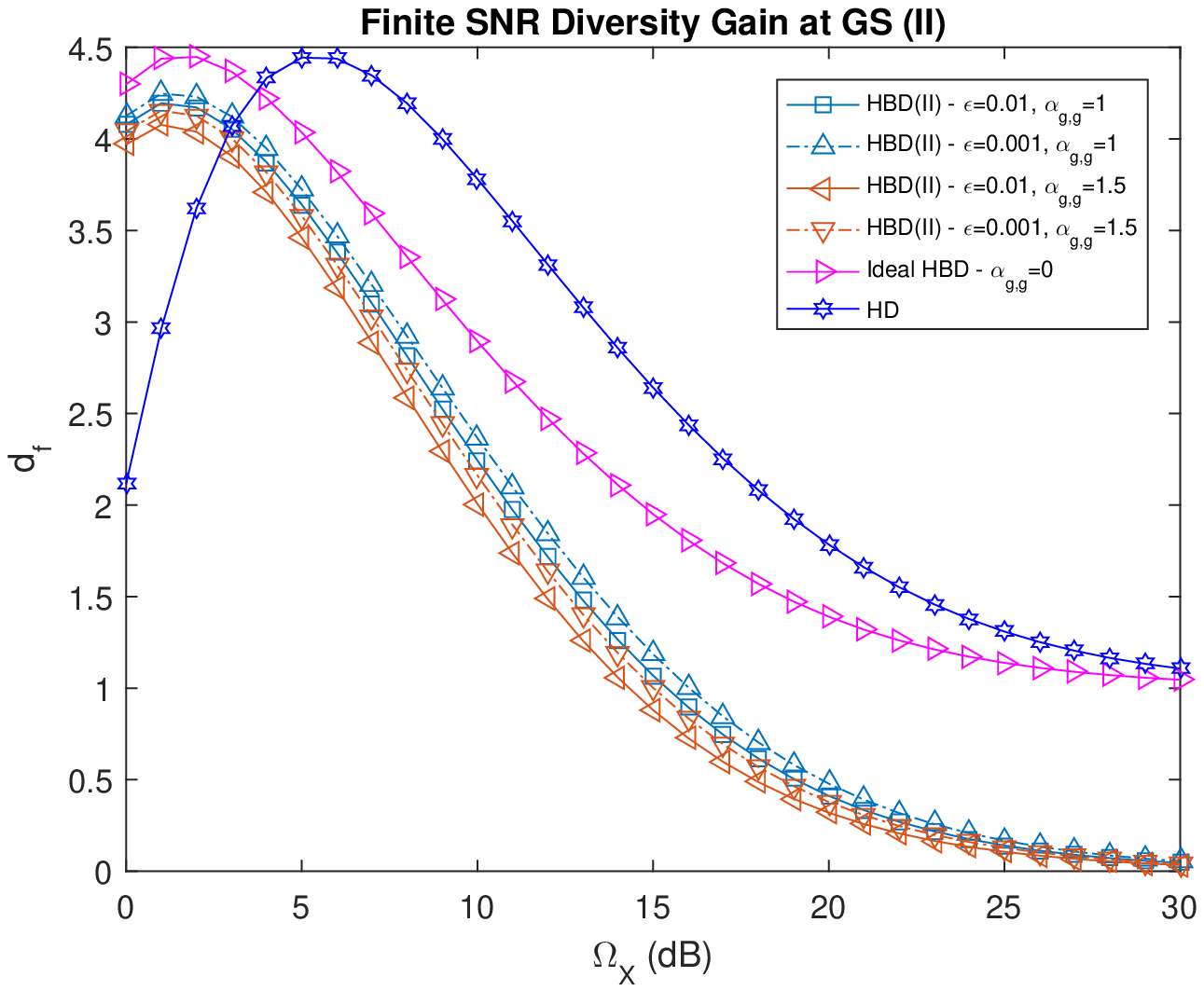}
\label{fig:fixed_df_gs}}
\caption{Outage probability and finite SNR diversity gain at GS (II detector) for phase noise strength $\gamma_{\phi}^2=-130$ dBm.}
\label{fig:gs_outage_diversity_gain}
\end{figure*}

The HBD outage probability at GS is shown in Fig. \ref{fig:fixed_pout_gs}. The ideal HBD and the HD outage probability are also plotted in Fig. \ref{fig:fixed_pout_gs} as a benchmark comparison. From Fig. \ref{fig:fixed_pout_gs}, it can be seen that $Pr\big(\mathcal{O}_{gs}^{HBD}\big)$ is close to the ideal HBD case, i.e., no interference, at low-to-moderate average received power ($\Omega_X$) and vice-versa. As expected, $Pr\big(\mathcal{O}_{gs}^{HBD}\big)$ is higher as SI channel estimation error ($\epsilon$) is increased. In addition, increasing the strength of the residual SI ($\alpha_{g,g}$) degrades the outage performance more than the increase in $\epsilon$ since a larger $\alpha_{g,g}$ corresponds to a higher average residual SI power, with phase noise ($\gamma^2_{\phi}$) scaled accordingly. In fact, $Pr\big(\mathcal{O}_{gs}^{HBD}\big)$ approaches the ideal HBD case when $\alpha_{g,g}=1, \epsilon=0.001$ at low $\Omega_X$ in Fig. \ref{fig:fixed_pout_gs}. Hence, sufficient SI mitigation is needed in order for the FD-enabled GS to outperform the HD-enabled GS. 

The finite SNR diversity gain at GS is shown in Fig. \ref{fig:fixed_df_gs}, where it can be seen that $d_{f,gs}^{HBD}$ peaks at $\Omega_X=2$ dB while $d_{f,gs}^{HD}$ peaks at $\Omega_X=6$ dB. \footnote{Higher diversity gain does not mean lower outage probability and vice-versa. To get a parametric representation for outage probability from diversity gain and SNR, the array gain, coding gain, or SNR offset, needs to be factored as shown in \cite{ordonez2012array} and the references therein. Similar analysis is needed from the interference-limited receiver's perspective to quantify the SNR offsets for different protocols at a given interference level, and it is left as a future extension of the current paper.} Additionally, (\ref{asymp_df_fixed_GS}) and (\ref{lim_df_fixed_hd}) are also confirmed in Fig. \ref{fig:fixed_df_gs} as $\Omega_X \to \infty$ and is also corroborated in Fig. \ref{fig:fixed_pout_gs}, where the slope of the outage probability curves become constant as $\Omega_X \to \infty$. In other words, the FD-enabled GS becomes interference-limited at high $\Omega_X$. Interestingly, in the absence of interference at the FD-enabled GS, $d_{f,gs}^{HBD} \to 1$ as $\Omega_X \to \infty$ since only SNR needs to be considered at GS. From Fig. \ref{fig:gs_outage_diversity_gain}, residual SI is the performance limiting factor for the FD-enabled GS. Therefore, it is important to sufficiently mitigate SI at each of the cascaded stages in Fig. \ref{fig:1} in order to keep the strength of the residual SI low for effective operation of the FD-enabled GS.

\begin{figure*}[]
\vspace{-0.5cm}
\centering
\subfloat[Outage probability comparison at AS-2]{\includegraphics [width=0.9\columnwidth]{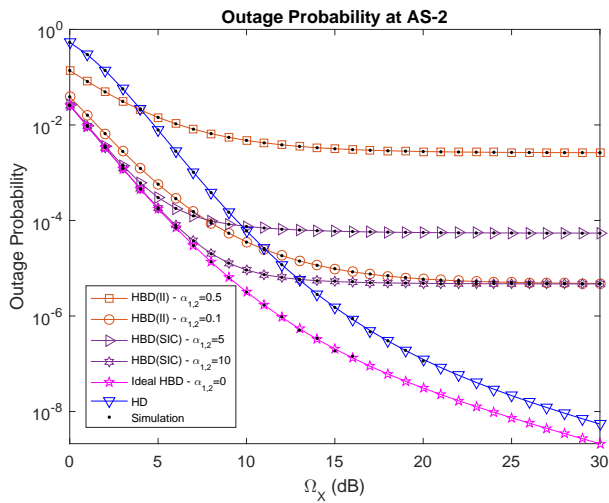}
\label{fig:fixed_pout_as2}}
\hfil
\subfloat[Finite SNR diversity gain comparison at AS-2]{\includegraphics [width=0.9\columnwidth]{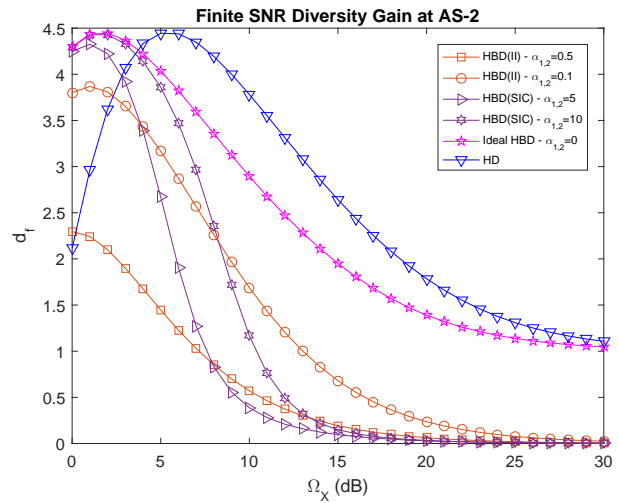}
\label{fig:fixed_df_as2}}
\caption{Outage probability and finite SNR diversity gain at AS-2 (II and SIC detectors) for $\alpha_{g,2}=1$, i.e., link between GS and AS-2 has same distance as the reference link ($d_{1,g}$).}
\label{fig_sim}
\end{figure*}

\subsubsection{\underline{\textit{Impact of Interference at AS-2}}}

\begin{observation}
\emph{The II and SIC detectors achieve lower outage probability and higher diversity gain than HD-mode at low SNR regimes and are interference-limited at high SNR regimes. For the SIC detector, strong interference at low SNR regime enables easy removal of the interfering signal.}
\end{observation}

The HBD outage probabilities at AS-2 for both II and SIC detectors are shown in Fig. \ref{fig:fixed_pout_as2}. It can be seen that the II detector at AS-2 outperforms the HD mode at low-to-moderate $\Omega_X$ when inter-AS interference ($\alpha_{1,2}$) is weak. The trend in Fig. \ref{fig:fixed_pout_as2} also suggests that the further reduction in $\alpha_{1,2}$ will enable the II detector at AS-2 to attain the ideal HBD outage performance for moderate $\Omega_X$, which is expected since $\alpha_{1,2}\to 0$ corresponds to diminishing levels of interference at AS-2. 

The SIC detector performs better than the HD mode at the low-to-moderate $\Omega_X$ when interference is strong, e.g., $\alpha_{1,2}=10$, since stage 1 of the SIC detector is more likely to detect and subtract $x_1[t]$. The resultant signal at stage 2 of the SIC detector is thus almost interference-free. As $\alpha_{1,2}$ increases, the SIC detector performance approaches that of the ideal HBD case due to the near perfect cancellation of interference in the first stage. When $\Omega_X>10$ dB for $\alpha_{1,2}\in\{5,10\}$, an error floor is present which verifies Corollary \ref{coro_asymp_df_fixed_AS2}. Similar error floor observations are also made for the II detector and it indicates that the II and SIC detectors become interference-limited at high $\Omega_X$. From a practical perspective, the trend in Fig. \ref{fig:fixed_pout_as2} shows that the II detector is well suited for en route scenarios with less congested flight routes such as those over sparsely populated or oceanic regions since the II detector experiences weak interference due to path loss as a result of large inter-aircraft or aircraft to GS distance. On the other hand, the SIC detector is suitable for use in congested airspace scenarios such as the landing or even continental en route scenarios as interference from nearby aircrafts can be effectively removed. Although HD-ACS has superior outage performance compared to the II and SIC detectors at high $\Omega_X$, the interference-limited HBD detectors can meet typical QoS requirements, e.g., frame error rate $\leq 10^{-3}$.

The finite SNR diversity gains, $d_{f,2}^{HBD(II)}$ and $d_{f,2}^{HBD(SIC)}$, at AS-2 are shown in Fig. \ref{fig:fixed_df_as2}. A trend similar to what was seen in Fig. \ref{fig:fixed_df_gs} can be found in Fig. \ref{fig:fixed_df_as2}, with $d_{f,2}^{HBD(II)}$ and $d_{f,2}^{HBD(SIC)}$ peaking at $\Omega_X=2$ dB, and $d_{f,2}^{HD}$ peaking at $\Omega_X=6$ dB. As expected, reducing $\alpha_{1,2}$ causes $d_{f,2}^{HBD(II)}$ to perform close to the ideal HBD case at low $\Omega_X$. Fig. \ref{fig:fixed_df_as2} also confirms (\ref{asymp_df_fixed_AS2}) for both the II and SIC detectors. It is clear that the SIC detector can attain an outage probability decay rate that is similar to the ideal HBD case when $\Omega_X\leq5$ dB. Further increasing $\alpha_{1,2}$ will enable $d_{f,2}^{HBD(SIC)}$ to be almost identical to the ideal HBD case at $\Omega_X\leq5$ dB since the system becomes noise-limited rather than interference-limited. The trends in Fig. \ref{fig:fixed_df_as2} are also reflected in Fig. \ref{fig:fixed_pout_as2} since the slope of the outage probability curves behave as indicated in (\ref{asymp_df_fixed_AS2}) and (\ref{lim_df_fixed_hd}) as $\Omega_X \to \infty$.

\begin{figure*}[]
\centering
\subfloat[System level outage probability]{\includegraphics [width=0.9\columnwidth]{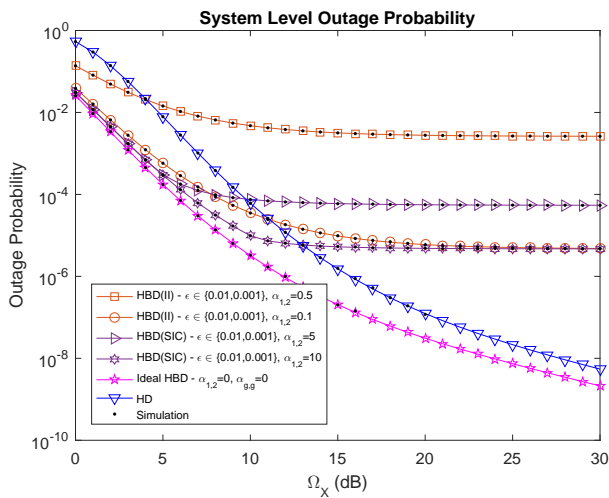}
\label{fig:fixed_pout_sys}}
\hfil
\subfloat[System level finite SNR diversity gain]{\includegraphics [width=0.9\columnwidth]{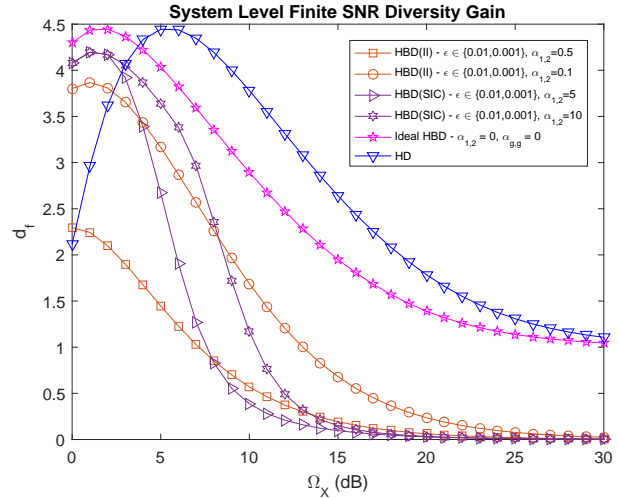}
\label{fig:fixed_df_sys}}
\caption{System level outage probability and finite SNR diversity gain (II and SIC detectors) for $\alpha_{g,2}=1$,$\alpha_{g,g}=1$,$\gamma_{\phi}^2=-130$ dBm, $\epsilon\in\{0.01, 0.001\}$.}
\label{fig_sim}
\end{figure*}

\subsubsection{\underline{\textit{Impact of Interference at System Level}}}

\begin{observation}
\emph{The system level performance of the HBD-ACS is constrained by inter-AS interference. When the II detector is considered, weak inter-AS interference enables near-ideal system level performance. Likewise for the SIC detector when strong inter-AS interference is present. If the SI suppression level is lower, then it is possible for the GS to be the bottleneck.}
\end{observation}

Fig. \ref{fig:fixed_pout_sys} and Fig. \ref{fig:fixed_df_sys} respectively shows the outage probability and finite SNR diversity gain at the system level. Through numerical analysis, we observed that $P_{out,system}^{HBD(II)}$ is dominated by the II detector at AS-2 for $\alpha_{1,2} \in \{0.1, 0.5\}$ and $0 \text{ dB} \leq \Omega_X \leq 30 \text{ dB}$, i.e., $Pr\big(\mathcal{O}_{gs}^{HBD}\big)<Pr\big(\mathcal{O}_{2}^{HBD(II)}\big)$ because inter-AS interference at AS-2 is stronger than the residual SI experienced at GS. Thus, although not shown in the figure, increasing $\alpha_{g,g}$ or $\epsilon$ does not affect $P_{out,system}^{HBD(II)}$ unless inter-AS interference is decreased. It can also be observed from Fig. \ref{fig:fixed_pout_sys} that $P_{out,system}^{HBD(II)} \leq P_{out,system}^{HD}$ when $\Omega_X\leq4$ dB, $\alpha_{1,2}=0.5$. When $\alpha_{1,2}=0.1$, $P_{out,system}^{HBD(II)} \leq P_{out,system}^{HD}$ for $\Omega_X\leq11$ dB. In fact, $P_{out,system}^{HBD(II)}$ approaches that of the ideal HBD case when $\alpha_{1,2}$ is decreased due to the near absence of inter-AS interference at the II detector and it also explains the trend seen in Fig. \ref{fig:fixed_df_sys} where it can be seen that $d_{f,system}^{HBD(II)}$ approaches that of the ideal HBD case when $\alpha_{1,2}$ is decreased. In other words, the decay of $P_{out,system}^{HBD(II)}$ approaches that of the ideal HBD case when inter-AS interference weakens, as reflected in Fig. \ref{fig:fixed_df_sys}, for $\Omega_X \leq 5$ dB. Therefore, when an II detector is used at AS-2, the inter-AS interference is the limiting factor for both $P_{out,system}^{HBD(II)}$ and $d_{f,system}^{HBD(II)}$.

When AS-2 adopts an SIC detector, $P_{out,system}^{HBD(SIC)}$ is dominated by GS when $\Omega_X\leq4$ dB and $\alpha_{1,2}=5$. Similar trends for the SIC detector are also seen in Fig. \ref{fig:fixed_df_sys} for $\Omega_X \leq 5$ dB. When $\Omega_X>4$ dB, $P_{out,system}^{HBD(SIC)}$ is dominated by AS-2 and it can be explained from the perspective of the two-stage SIC detector at AS-2. When $\alpha_{1,2}=5$, $x_1[t]$ is five times stronger than the SOI from GS ($x_{gs}[t]$). In addition, at stage 1 of the SIC detector, noise power ($\sigma^2_{2}$) is stronger than $x_{gs}[t]$ when $\Omega_X\leq4$ dB. Thus, the SIC detector is more likely to detect and cancel $x_1[t]$ which results in $Pr\big(\mathcal{O}_{gs}^{HBD}\big)>Pr\big(\mathcal{O}_{2}^{HBD(SIC)}\big)$ due to residual SI at GS. When $\Omega_X>4$ dB, $\sigma_2^2$ will be weaker than $x_{gs}[t]$ at stage 1 of the SIC detector. Consequently, the SIC detector is less likely to detect and cancel $x_1[t]$, leading to $Pr\big(\mathcal{O}_{gs}^{HBD}\big)<Pr\big(\mathcal{O}_{2}^{HBD(SIC)}\big)$. When $\alpha_{1,2}=10$, $P_{out,system}^{HBD(SIC)}$ is dominated by GS for $\Omega_X \leq 10$ dB due to stronger interference at AS-2, with $P_{out,system}^{HBD(SIC)}$ close to that of the ideal HBD case. Further increasing $\alpha_{1,2}$ enables $P_{out,system}^{HBD(SIC)}$ to reach near-ideal HBD performance for a wider $\Omega_X$ range due to the increased likelihood of successfully detecting and canceling $x_1[t]$, thus explaining the trend in Fig. \ref{fig:fixed_df_sys}. Hence, the strength of the interference from AS-1 ($\alpha_{1,2}$) is the main limiting factor for both $P_{out,system}^{HBD(SIC)}$ and $d_{f,system}^{HBD(SIC)}$ when a SIC detector is used at AS-2. 


From Fig. \ref{fig:fixed_pout_sys} and Fig. \ref{fig:fixed_df_sys}, the outage and finite SNR diversity gain analysis has highlighted the feasibility of HBD-ACS over legacy HD-ACS in weak and strong interference scenarios through the II and SIC detectors, respectively. For instance, weak and strong interference scenarios could involve en route flights over sparely and densely populated airspace, respectively. From the aeronautical perspective, the proposed HBD-ACS has better reliability over HD-ACS while providing more throughput than legacy HD systems. 

\subsection{Finite SNR DMT Analysis}

\subsubsection{\underline{\textit{Impact of Residual SI at GS}}}

\begin{observation}
\emph{The FD-enabled GS achieves non-zero diversity gain for a larger range of multiplexing gains compared to the HD GS.}
\end{observation}

\begin{figure*}[]
\centering
\subfloat[GS (II detector).]{\includegraphics [width=0.9\columnwidth]{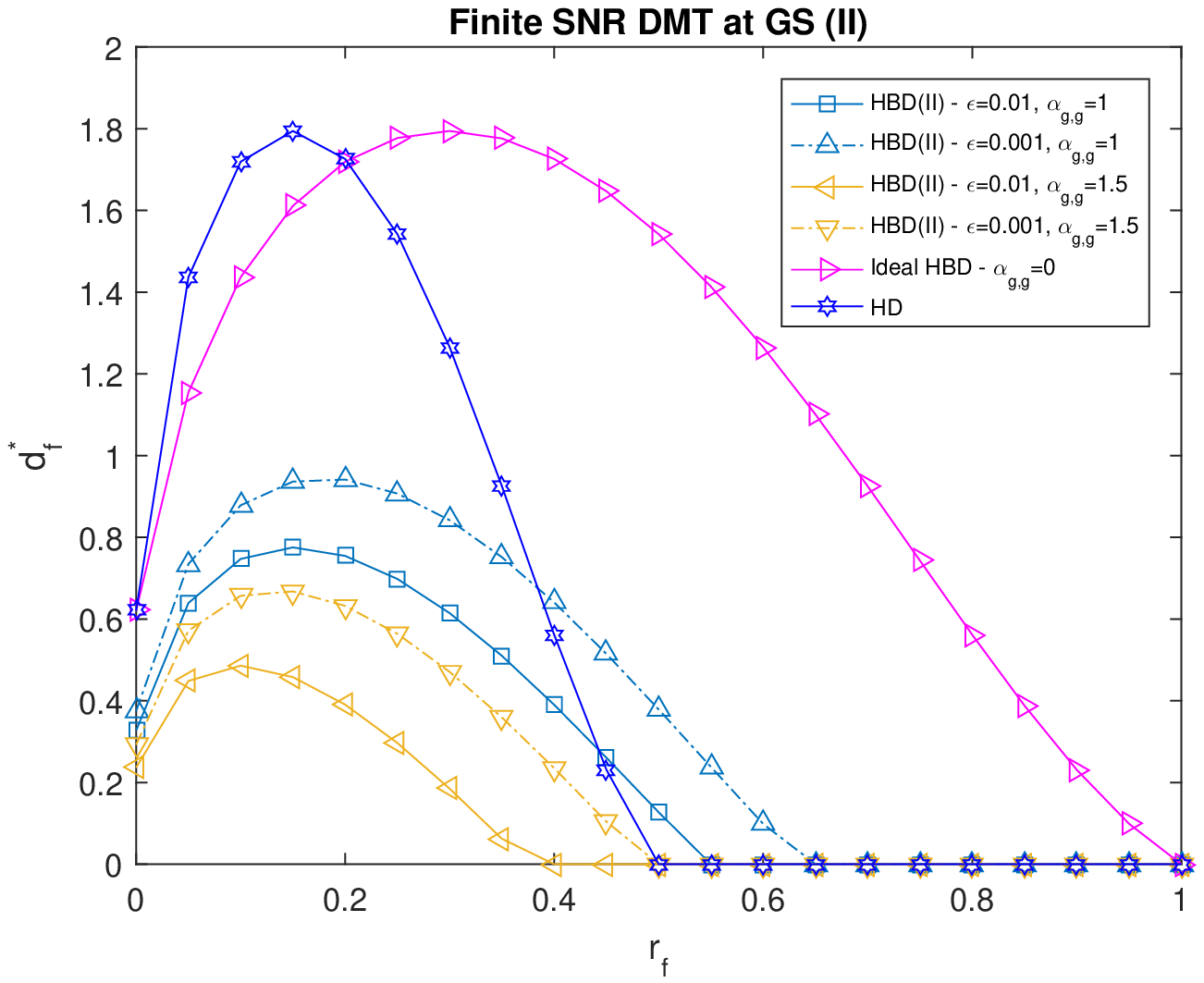}
\label{fig:var_df_gs}}
\hfil
\subfloat[AS-2 (II and SIC detector).]{\includegraphics [width=0.9\columnwidth]{var_df_as2.eps}
\label{fig:var_df_as2}}
\caption{Finite SNR DMT for $\alpha_{g,2}=1, \gamma_{\phi}^2=-130 \text{ dBm}, \Omega_X=10 \text{ dB}$.}
\label{fig_sim}
\end{figure*}

Fig. \ref{fig:var_df_gs} shows the finite SNR diversity gain at GS, where it is evident that the stronger residual SI due to SI channel estimation error ($\epsilon$) or phase noise ($\gamma_{\phi}^2$) reduces $d_{f,gs}^{HBD*}$. Increasing the strength of the residual SI ($\alpha_{g,g}$) affects $d_{f,gs}^{HBD*}$ more than increasing the SI channel estimation error ($\epsilon$) since the effect of phase noise ($\gamma_{\phi}^2$) on the residual SI is amplified. From the outage probability perspective, increasing residual SI results in a slower decay rate of the outage probability, which lowers $d_{f,gs}^{HBD*}$. However, it does not imply that outage probability is better when a higher maximum value for $d_{f,gs}^{HBD*}$ is attained. Nonetheless, the range of $r_f$ for which $d_{f,gs}^{HBD*}\geq d_{f,gs}^{HD*}$ increases when the strength of the residual SI ($\alpha_{g,g}$) decreases and vice versa. Therefore, FD-enabled GS can experience improved DMT as residual SI decreases, which is evident in Fig. \ref{fig:var_df_gs} for $\alpha_{g,g}=1$. Although $d_{f,gs}^{HBD*}$ is limited by residual SI, the importance of proper SI mitigation is again emphasized since it is still feasible for GS to be FD-enabled if operating at a higher $r_f$ is the objective of an ACS.

\subsubsection{\underline{\textit{Impact of Interference at AS-2}}}

\begin{observation}
\emph{At low multiplexing gains, the II and SIC detectors have lower finite SNR diversity gain. In contrast, at high multiplexing gains, the II and SIC detectors achieve near-ideal finite SNR diversity gain under weak and strong inter-AS interference, respectively.}
\end{observation}

Fig. \ref{fig:var_df_as2} shows the finite SNR diversity gain at AS-2. The trends seen in Fig. \ref{fig:var_df_as2} are similar to what was seen in \cite[Fig. 4]{narasimhan2006finite}, with lower $d_{f,2}^{HBD(i)*}, i \in \{II,SIC\}$ and $d_{f,2}^{HD*}$ observed as $r_f \to 0$. It has been pointed out by Narasimhan \cite{narasimhan2006finite} and Shin et al. \cite{shin2008diversity} that Rician fading outage probability curves are influenced by Rician $K$ factors. In particular, increasing the Rician $K$ factor causes the slope of outage probability curves to become steeper \cite[Fig. 2]{shin2008diversity}. From a finite SNR DMT perspective, $r_f \to 0$ causes $K_{X_{gs}}$ to have less impact on the outage performance at AS-2. 

On the other hand, Fig. \ref{fig:var_df_as2} also suggests that the II and SIC detectors are able to provide better reliability at higher multiplexing gains compare to HD systems. At high multiplexing gains, if the inter-AS interference reduces, then $d_{f,2}^{HBD(II)*} \geq d_{f,2}^{HD*}$. On the other hand, at low multiplexing gains, $d_{f,2}^{HBD(II)*} < d_{f,2}^{HD*}$ even at low inter-AS interference. In fact, $d_{f,2}^{HBD(II)*}$ approaches that of the ideal HBD case as $\alpha_{1,2} \to 0$ since the signal at the II detector is almost interference-free. As a consequence, the resultant outage probability decay rate becomes similar to that of the ideal HBD case. When a SIC detector is adopted at AS-2, $d_{f,2}^{HBD(SIC)*} \geq d_{f,2}^{HD*}$ as inter-AS interference increases (for example, refer to $d_{f,2}^{HBD(SIC)*}$ at $\alpha_{1,2} = 14.3$ in Fig. \ref{fig:var_df_as2}). As $\alpha_{1,2}\to\infty$, it becomes easier to detect and remove $x_1[t]$ at the two-stage SIC detector. When coupled with the lower threshold requirement of the SIC detector, as compared to HD systems, the SIC detector can potentially achieve superior diversity gains over HD systems in strong interference scenarios. Moreover, at large values of $\alpha_{1,2}$, if the multiplexing gain is high, the achievable $d_{f,2}^{HBD(SIC)*}$ matches the ideal HBD case. As shown in Fig. \ref{fig:var_df_as2}, at low multiplexing gain, the achievable $d_{f,2}^{HBD(SIC)*}$ is close to that of the ideal HBD case. Therefore, the II and SIC detectors provides better reliability at higher multiplexing gains compared to HD-ACS in the presence of weak and strong interference, respectively. However, at low multiplexing gains, HD-ACS exhibited better reliability than the II and SIC detectors.

\begin{figure} []
\centering
\includegraphics [width=0.9\columnwidth]{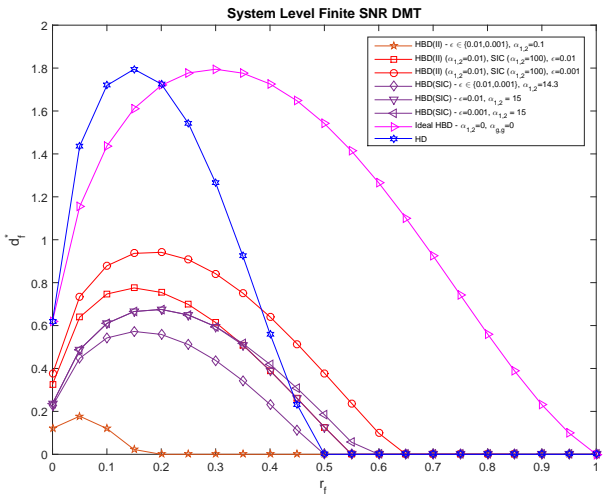} 
\caption{System level finite SNR DMT (II and SIC detectors) for $\alpha_{g,2}=1$, $\alpha_{g,g}=1$, $\gamma_{\phi}^2=-130 \text{ dBm}$, $\Omega_X=10 \text{ dB}$.}
\label{fig:var_df_sys}
\end{figure}

\subsubsection{\underline{\textit{Impact of Interference at System Level}}}

\begin{observation}
\emph{At high multiplexing gains, the HBD-ACS achieves better finite SNR diversity gain at the system level than the HD-ACS and is also constrained by inter-AS interference and residual SI.}
\end{observation}

Fig. \ref{fig:var_df_sys} shows the system level finite SNR diversity gain for HBD-ACS ($d_{f,system}^{\beta*}$) and HD-ACS ($d_{f,system}^{HD*}$) for $\beta \in \{HBD(II), HBD(SIC)\}$. From Fig. \ref{fig:var_df_sys}, it is evident that $d_{f,system}^{HBD(II)*} > d_{f,system}^{HD*}$ and $d_{f,system}^{HBD(SIC)*} > d_{f,system}^{HD*}$ as $r_f$ increases, and it enables an HBD-ACS to provide better reliability at higher multiplexing gain than HD-ACS since HBD-ACS requires a lower operating threshold than existing HD-ACS at both GS and AS-2. However, the degree of improvement that HBD-ACS has over HD-ACS is constrained by the strength of interference experienced at GS and AS-2 in the HBD-ACS.

When the II detector is adopted at AS-2 for weak interference scenarios, $d_{f,gs}^{HBD*} > d_{f,2}^{HBD(II)*}$ for $\alpha_{1,2}=0.1$. Reducing the strength of the inter-AS interference ($\alpha_{1,2}=0.01$) causes $d_{f,2}^{HBD(II)*} > d_{f,gs}^{HBD*}$, with lower SI channel estimation error ($\epsilon $) corresponding to higher $d_{f,system}^{HBD(II)*}$. In the presence of strong interference at AS-2 ($\alpha_{1,2}=100$), adopting the SIC detector at AS-2 results in $d_{f,2}^{HBD(SIC)*} > d_{f,gs}^{HBD*}$. However, when interference from AS-1 is not as strong, e.g., $\alpha_{1,2} \in \{14.3, 15\}$, then $d_{f,gs}^{HBD*} > d_{f,2}^{HBD(SIC)*}$. From Fig. \ref{fig:var_df_sys}, the reliability of the HBD-ACS depends on the inter-AS interference at AS-2 for both II and SIC detectors and residual SI at GS. Furthermore, it is possible for the proposed HBD-ACS to attain finite SNR DMT curves that are identical to the ideal HBD case at sufficiently low residual SI.

From Fig. \ref{fig:var_df_sys}, the trends show that the proposed HBD-ACS is a viable alternative to legacy HD-ACS in weak and strong interference scenarios. In particular, the proposed HBD-ACS can operate at a higher multiplexing gain than legacy HD-ACS, thus offering better throughput and reliability compared to the latter.


\section{Conclusion}
An HBD-ACS consisting of an FD-enabled GS and two HD ASs simultaneously communicating on the same spectrum is proposed to improve spectrum utilization. To investigate the impact of interference on the proposed HBD-ACS, closed-form outage probability and finite SNR diversity gain expressions are presented in this paper for a SIC detector over Rician fading aeronautical channels. Through outage and finite SNR diversity gain analysis, it is established that residual SI is the main limiting factor at the FD-enabled GS. Therefore, the need for sufficient SI mitigation must be properly addressed in a HBD-ACS. At AS-2, inter-AS interference is the main limiting factor for both II and SIC detectors. At the system level, the proposed HBD-ACS is found to be very suitable for weak and strong interference scenarios for the II and SIC detectors, respectively. The proposed HBD-ACS is also able to achieve superior outage performance and better diversity gains at low-to-moderate SNRs compared to existing HD-ACS for both weak and strong interference scenarios. Finite SNR DMT analysis has also revealed that HBD-ACS can achieve interference-free diversity gain if residual SI is sufficiently suppressed, enabling HBD-ACS to be more reliable than HD-ACS at higher multiplexing gains while operating at low SNR ranges.


\section*{Acknowledgment}
This research is jointly funded by Airbus Singapore Pte Ltd and the Singapore Economic Development Board (EDB). The authors would also like to thank the editor and the anonymous reviewers whose feedback helped us to improve the quality of this work.

\appendices 
\section{Proof of (\ref{P_out_as2_SIC})} \label{SIC_proof}
Let $X_{gs}$ be the average received power of the SOI with non-centered chi-squared probability density function (PDF) $f_{X_{gs}}(x) = \frac{K_{X_{gs}}+1}{\Omega_{X}\alpha_{g,2}} \exp\left(-K_{X_{gs}}-\frac{K_{X_{gs}} + 1}{\Omega_{X}\alpha_{g,2}}x\right) I_{0}\left(2\sqrt{\frac{K_{X_{gs}}(K_{X_{gs}}+1)}{\Omega_{X}\alpha_{g,2}}x}\right)$, where $I_{0}\left(\cdot\right)$ is the modified Bessel function of the first kind with zero order \cite[Eq. (8.445)]{gradshteyn2014table}. Similarly, let $Y_1$ be the average received power of the interfering signal with non-centered chi-squared PDF $f_{Y_1}(y) = \frac{K_{Y_1}+1}{\Omega_{X}\alpha_{1,2}}\exp\left(-K_{Y_1}-\frac{K_{Y_1} + 1}{\Omega_{X}\alpha_{1,2}}y\right) I_{0}\left(2\sqrt{\frac{K_{Y_1}(K_{Y_1}+1)}{\Omega_{X}\alpha_{1,2}}y}\right)$. 

The closed-form SIC outage probability at AS-2 is equivalent to computing the sum of the areas of outage regions $P_1$ and $P_2$, i.e.,$Pr(\mathcal{O}_{2}^{HBD(SIC)}) = P_1 + P_2$. Let the outage regions be defined as $P_1 = Pr\left\{Y_1 < \gamma_{th,gs}^{HBD}\left(1+X_{gs}\right) \right\}$ and $P_2 = Pr\left\{Y_1 \geq \gamma_{th,2}^{HBD}(1+X_{gs}), X_{gs} < \gamma_{th,2}^{HBD} \right\}$. The expression for $P_1$ can be rewritten as \cite{rached2017unified}:
\begin{eqnarray} \label{eq_P1_2}
P_1 & \hspace{-0.25cm} = & \hspace{-0.25cm} \int_{0}^{\infty} \int_{0}^{\gamma_{th,gs}^{HBD}(1+X_{gs})} f_{Y_1}(y) f_{X_{gs}}(x) dydx \nonumber \\
 & \hspace{-0.25cm} = & \hspace{-0.25cm} \sum_{q\geq0}\sum_{l=0}^{q+1}\alpha(q,\Omega_X\alpha_{1,2},K_{Y_{1}},\gamma_{th,gs}^{HBD})\binom{q+1}{l}E\{X_{gs}^l\}_,
\end{eqnarray}
where $E\{\cdot\}$ represents the expectation function. The expression for $P_2$ can be expressed as:
\begin{eqnarray} \label{eq_P2_2}
P_2 & = & \int_{0}^{\gamma_{th,2}^{HBD}} \int_{\gamma_{th,gs}^{HBD}(1+X_{gs})}^{\infty} f_{Y_1}(y) f_{X_{gs}}(x) dydx \nonumber\\
 & = & \int_{0}^{\gamma_{th,2}^{HBD}} \hspace{-0.2cm} Q_1\left( \sqrt{2K_{Y_1}}, \sqrt{\frac{2(K_{Y_1}+1)\gamma_{th,gs}^{HBD}(1+x)}{\Omega_{X}\alpha_{1,2}}} \right) \nonumber \\
 & & \hspace{4.7cm} \times f_{X_{gs}}(x) dx_. 
\end{eqnarray}

From \cite{andras2011generalized}, $f_{X_{gs}}(x) = \sum_{j\geq0}\alpha(j,\Omega_X\alpha_{g,2},K_{X_{gs}},1)x^j$. Thus, (\ref{eq_P2_2}) can be rewritten as:
\begin{eqnarray} \label{eq_P2_3}
P_2 & = & 1 - Q_1\left( \sqrt{2K_{X_{gs}}}, \sqrt{\frac{2(K_{X_{gs}}+1)\gamma_{th,2}^{HBD}}{\Omega_{X}\alpha_{g,2}}} \right) \nonumber \\
 & & \hspace{-1.3cm} - \int_{0}^{\gamma_{th,2}^{HBD}} \left(\sum_{j\geq0}\alpha(j,\Omega_X\alpha_{g,2},K_{X_{gs}},1)x^j\right) \nonumber\\
 & & \hspace{-0.7cm} \times \Bigg(\sum_{n\geq0}\alpha(n,\Omega_X\alpha_{1,2},K_{Y_1},\gamma_{th,gs}^{HBD})\sum_{i=0}^{n+1}\binom{n+1}{i}x^{i}\Bigg) dx_.
\end{eqnarray}

Let $c(n) = \alpha(n,\Omega_X\alpha_{1,2},K_{Y_1},\gamma_{th,gs}^{HBD})\sum_{i=0}^{n+1}\binom{n+1}{i}x^{i}$ and $ d(j) = \alpha(j,\Omega_X\alpha_{g,2},K_{X_{gs}},1)x^j$, then the integral in (\ref{eq_P2_3}) can be written as \cite{andras2011generalized, bartoszewicz2012algebrability}:
\begin{eqnarray} \label{eq_P2_4}
 & & \hspace{-0.5cm} \int_{0}^{\gamma_{th,2}^{HBD}}\left(\sum_{n\geq0}c(n)\right)\left(\sum_{j\geq0}d(j)\right) dx \nonumber \\
 & & = \int_{0}^{\gamma_{th,2}^{HBD}} \sum_{n\geq0}\sum_{i=0}^{n}c(i)d(n-i) dx \nonumber\\
 & & = \sum_{n\geq0}\sum_{i=0}^{n}\alpha(i,\Omega_X\alpha_{1,2},K_{Y_1},\gamma_{th,gs}^{HBD}) \nonumber \\
 & & \hspace{0.25cm} \times \alpha(n-i,\Omega_X\alpha_{g,2},K_{X_{gs}},1) \sum_{j=0}^{i+1}\binom{i+1}{j}\int_{0}^{\gamma_{th,2}^{HBD}}x^{j+n-i}dx \nonumber\\
 & & = \sum_{n\geq0}\sum_{i=0}^{n}\sum_{j=0}^{i+1} \alpha(i,\Omega_X\alpha_{1,2},K_{Y_1},\gamma_{th,gs}^{HBD}) \nonumber \\
 & & \hspace{0.4cm} \times \alpha(n-i,\Omega_X\alpha_{g,2},K_{X_{gs}},1)\binom{i+1}{j}\frac{(\gamma_{th,2}^{HBD})^{j+n-i+1}}{j+n-i+1}_.
\end{eqnarray}

Combining (\ref{eq_P1_2}) and (\ref{eq_P2_4}), the expression in (\ref{P_out_as2_SIC}) can be obtained.

In (\ref{eq_P2_2}), $Q_1\left( \sqrt{2K_{Y_1}}, \sqrt{\frac{2(K_{Y_1}+1)\gamma_{th,gs}^{HBD}(1+x)}{\Omega_{X}\alpha_{1,2}}} \right) = 1 - \sum_{n\geq0}{\alpha(n,\Omega_X\alpha_{1,2},K_{Y_1},\gamma_{th,gs}^{HBD})(1+x)^{n+1}}$ if $\frac{K_{Y_1}+1}{\Omega_{X}\alpha_{1,2}}(\gamma_{th,gs}^{HBD})(1+x)\geq0$ \cite{andras2011generalized}. In addition, the PDF $f_{X_{gs}}(x)$ can be expressed as a convergent power series if $\frac{K_{X_{gs}}+1}{\Omega_{X}\alpha_{g,2}}x\geq0$ \cite{andras2011generalized}. Assuming the power series in (\ref{eq_P2_3}) is convergent, the resultant product of the power series in (\ref{eq_P2_4}) will also be convergent \cite{bartoszewicz2012algebrability}. Similarly in (\ref{eq_P1_2}), the power series is convergent if $\gamma_{th,gs}^{HBD}\leq \frac{(\Omega_{X}\alpha_{1,2})/(1+K_{Y_1})}{2(\Omega_{X}\alpha_{g,2})/(1+K_{X_{gs}})}$ \cite{rached2017unified}. Therefore, the closed-form expression in (\ref{P_out_as2_SIC}) holds if the power series in (\ref{eq_P1_2}) and (\ref{eq_P2_4}) are convergent. This completes the proof.

\section{Proof of Corollary \ref{coro_asymp_df_fixed_GS}} \label{coro_asymp_df_fixed_GS_proof}
From (\ref{P_out_gs_II}) and (\ref{df_fixed_GS}), $\left(\Omega_X\right)^{l_1+l_2-q-1} < 1$ when $l_1 + l_2 + l_3 \leq q$. Thus, $\lim_{\Omega_X\to\infty} \left(\Omega_X\right)^{l_1+l_2-q-1} \\ = 0, l_1 + l_2 + l_3 \leq q$. Therefore, only $l_1 + l_2  + l_3 = q + 1$ needs to be considered, which consequently leads to the numerator in (\ref{df_fixed_GS}) to be zero, i.e., $l_1 + l_2 - q - 1 = 0$. This completes the proof.

\section{Proof of Corollary \ref{coro_asymp_df_fixed_AS2}} \label{coro_asymp_df_fixed_AS2_proof}
To evaluate $\lim_{\Omega_X\to\infty} d_{f,2}^{HBD(II)}$, the approach seen in (\ref{asymp_df_fixed_GS}) can be used. Starting with the denominator of $d_{f,2}^{HBD(II)}$, $\left(\Omega_X\right)^{l-q-1} < 1$ when $l \leq q$. Thus, $\lim_{\Omega_X\to\infty} \left(\Omega_X\right)^{l-q-1} = 0$ when $l \leq q$. In the numerator, $(l-q-1) \left(\Omega_X\right)^{l-q-2} = 0$ when $l=q+1$. Similarly, to evaluate $\lim_{\Omega_X\to\infty} d_{f,2}^{HBD(SIC)}$, we first begin with the denominator of $d_{f,2}^{HBD(SIC)}$. Specifically, $\lim_{\Omega_X\to\infty}(\Omega_X)^{l-q-1}=0 $ when $l\leq{q}$ and $(\Omega_X)^{l-q-1}=1$ when $l=q+1$. For $(\Omega_X)^{-m-1}$, $\lim_{\Omega_X\to\infty}(\Omega_X)^{-m-1}=0 $ for $m\geq0$ and for $(\Omega_X)^{-n-2}$, $\lim_{\Omega_X\to\infty}(\Omega_X)^{-n-2}=0 $ for $n\geq0$. In the numerator, $(l-q-1) \left(\Omega_X\right)^{l-q-2} = 0$ when $l=q+1$. Additionally, $\lim_{\Omega_X\to\infty} (\Omega_X)^{-m-2}=0$ when $m\geq0$ and $\lim_{\Omega_X\to\infty} (\Omega_X)^{-n-3}=0$ when $n\geq0$. This completes the proof.

\section{Proof of Corollary \ref{coro_lim_df_fixed_hd}} \label{coro_lim_df_fixed_hd_proof}
At GS, the asymptotic behavior of $d_{f,gs}^{HD}$ can be easily evaluated after some simplifications as shown below:
\begin{eqnarray} \label{lim_df_fixed_hd_GS}
 & & \hspace{-1.4cm} \lim_{\Omega_X \to \infty} d_{f,gs}^{HD} \nonumber \\
 & \hspace{-1.1cm} = & \hspace{-0.7cm} \lim_{\Omega_X \to \infty} \frac{ -\sum_{m\geq0} \alpha\big(m,1,K_{X_1},\gamma_{th,gs}^{HD}\big) (-m-1) (\Omega_{X})^{-m} }{\sum_{m\geq0} \alpha\big(m,1,K_{X_1},\gamma_{th,gs}^{HD}\big) (\Omega_{X})^{-m} }_. 
\end{eqnarray}

From (\ref{lim_df_fixed_hd_GS}), It can be seen that $\lim_{\Omega_X\to\infty} (\Omega_{X})^{-m}=1 $ when $m=0$, and $\lim_{\Omega_X\to\infty} (\Omega_{X})^{-m} = 0 $ when $m>0$. Thus, when evaluating (\ref{lim_df_fixed_hd_GS}), only $m=0$ needs to be considered. The asymptotic behavior of $d_{f,2}^{HD}$ can also be proven using the same approach. This completes the proof.

\ifCLASSOPTIONcaptionsoff
  \newpage
\fi
\bibliographystyle{IEEEtran} 
\bibliography{IEEEabrv,ref}

\end{document}